\documentclass[sigconf]{acmart}

\usepackage{seqsplit}

\AtBeginDocument{%
  \providecommand\BibTeX{{%
    \normalfont B\kern-0.5em{\scshape i\kern-0.25em b}\kern-0.8em\TeX}}}

\setcopyright{acmcopyright}
\copyrightyear{2022}
\acmYear{2022}
\acmDOI{XXXXXXX.XXXXXXX}

\acmBooktitle{} 
\acmPrice{15.00}
\acmISBN{978-1-4503-XXXX-X/18/06}
\usepackage{algorithm}
\usepackage[noend]{algpseudocode}
\usepackage{mathtools}
\usepackage{xcolor}
\definecolor{redviolet}{HTML}{C71585}
\definecolor{darkyellow}{HTML}{F6BE00}
\usepackage{subcaption}
\usepackage{array}

\usepackage{colortbl}

\copyrightyear{2023} 
\acmYear{2023} 
\setcopyright{acmlicensed}\acmConference[CHI '23]{Proceedings of the 2023 CHI Conference on Human Factors in Computing Systems}{April 23--28, 2023}{Hamburg, Germany}
\acmBooktitle{Proceedings of the 2023 CHI Conference on Human Factors in Computing Systems (CHI '23), April 23--28, 2023, Hamburg, Germany}
\acmPrice{15.00}
\acmDOI{10.1145/3544548.3580644}
\acmISBN{978-1-4503-9421-5/23/04}

\begin{document}


\title[Upvotes? Downvotes? No Votes?]{\textcolor{black}{Upvotes? Downvotes? No Votes? Understanding the relationship between reaction mechanisms and political discourse on Reddit}}

\author{Orestis Papakyriakopoulos}
\authornote{Most work was done during author's research appointment at the Princeton Center for Information Technology Policy}


\affiliation{%
  \institution{Sony AI}
  \city{Zurich}
  \state{}
  \country{Switzerland}
}
\email{orestis.papakyriakopoulos@sony.com}

\author{Severin Engelmann}
\affiliation{%
  \institution{Technical University of Munich}
  \streetaddress{}
  \city{Munich}
  \country{Germany}}
\email{severin.engelmann@tum.de}

\author{Amy Winecoff}
\affiliation{%
  \institution{Princeton University}
  \city{Princeton}
  \state{New Jersey}
  \country{USA}
}
\email{aw0934@princeton.edu}

\renewcommand{\shortauthors}{Papakyriakopoulos, Engelmann \& Winecoff}

\begin{abstract}
A significant share of political discourse occurs online on social media platforms. Policymakers and researchers try to understand the role of social media design in shaping the quality of political discourse around the globe. In the past decades, scholarship on political discourse theory has produced distinct characteristics of different types of prominent political rhetoric such as deliberative, civic, or demagogic discourse. This study investigates the relationship between social media reaction mechanisms (i.e., upvotes, downvotes) and political rhetoric in user discussions by engaging in an in-depth conceptual analysis of political discourse theory. First, we analyze 155 million user comments in 55 political subforums on Reddit between 2010 and 2018 to explore whether users' style of political discussion aligns with the essential components of deliberative, civic, and demagogic discourse. Second, we perform a \textcolor{black}{quantitative study} that combines confirmatory factor analysis with difference in differences models to explore whether different reaction mechanism schemes (e.g., upvotes only, upvotes and downvotes, no reaction mechanisms) correspond with political user discussion that is more or less characteristic of deliberative, civic, or demagogic discourse. We produce three main takeaways. First, despite being ``ideal constructs of political rhetoric,'' we find that political discourse theories describe political discussions on Reddit to a large extent. Second, we find that discussions in subforums with only upvotes, or both up- and downvotes \textcolor{black}{are associated with} user discourse that is more deliberate and civic. Third, and perhaps most strikingly, social media discussions are most demagogic in subreddits with no reaction mechanisms at all. These findings offer valuable contributions for ongoing \textcolor{black}{policy discussions on the relationship between social media interface design and respectful political discussion among users}.\footnote{Our source code is available for public use under \url{https://github.com/civicmachines/political_discourse_reaction_design}}
\end{abstract}



\maketitle

\section{Introduction}

\textcolor{black}{Political exchange among citizens occurs largely on social media platforms.} Platforms have become the ``de facto public sphere''  ~\cite{tufekci2017twitter} to discuss political topics  ~\cite{stier2018election}, perform political campaigning  ~\cite{kreiss2018technology}, and communicate important messages for the pursuit of social causes and protests  ~\cite{jackson2020hashtagactivism}. However, they have also become common places for users to engage in hateful  ~\cite{mathew2019spread} and low-credibility political rhetoric  ~\cite{vosoughi2018spread}. Social media platforms are not simply  digital representations of offline political activity.  They are key spaces for articulation, organization, and implementation of political action  ~\cite{burrell2021society}. Indeed, they are not intermediaries of communication processes, but function as their \textit{curators}  ~\cite{gillespie2017platforms}. When billions of people interact in a common space, a platform's design has enormous power over the production, mediation, and dissemination of user discussions. A platform's choice of communication  reaction mechanisms (i.e., ``likes,'' ``upvotes/downvotes'' etc.) and its recommendation algorithms critically influence the nature of user interactions on the platform. On the one hand,  such reaction mechanisms -- as means of evaluation -- condition what type of information users produce, and, on the other hand, recommendation algorithms determine what information users will come to interact with  ~\cite{papacharissi2010networked, lazer2015rise, bond201261, meta_2021}. 

Prior research studies illustrate that social media platforms have transformative effects on political communication, albeit intense debates about whether and how design features influence the quality of discourse among users  ~\cite{tucker2018social,tucker2017liberation,margetts2018rethinking,bail2018exposure,barbera2014social}. Scholarship on political rhetoric has proposed a vast set of rhetoric devices that are essential to the corresponding modes of political discourse  ~\cite{ayers2013comparing,monnoyer2012technology,engesser2017populism,kushin2009getting,jennings2021social}. 

In this research, we \textit{first} explore to what extent essential  components of political discourse theories characterize political discussions on Reddit. \textit{Second}, we study how specific digital  reaction mechanisms on platforms (such as liking, voting, retweeting) that navigate user feedback on content  ~\cite{dellarocas2003digitization} and optimize platforms' recommendation systems  ~\cite{covington2016deep,tiktok_2020} impact the prevalence of specific types of political discourse  ~\cite{halpern2013social}. To this end, we analyze an extensive dataset of political discussions on Reddit against the theoretical framework of three prominent political discourse theories: deliberative, civic, and demagogic discourse. Finally, we answer the following research questions.

\begin{description}
\item[RQ1:] To what extent can essential rhetoric components of deliberative, civic, and demagogic discourse describe users' political discussions on Reddit?

\vspace{2mm}

\item[RQ2:] Are different reaction mechanisms (i.e., upvotes, downvotes, no votes) associated with  different rhetoric components of political discourse in political discussions on Reddit?
\end{description}

\begin{figure*}[!h]
    \centering
    \includegraphics[width=0.9\textwidth]{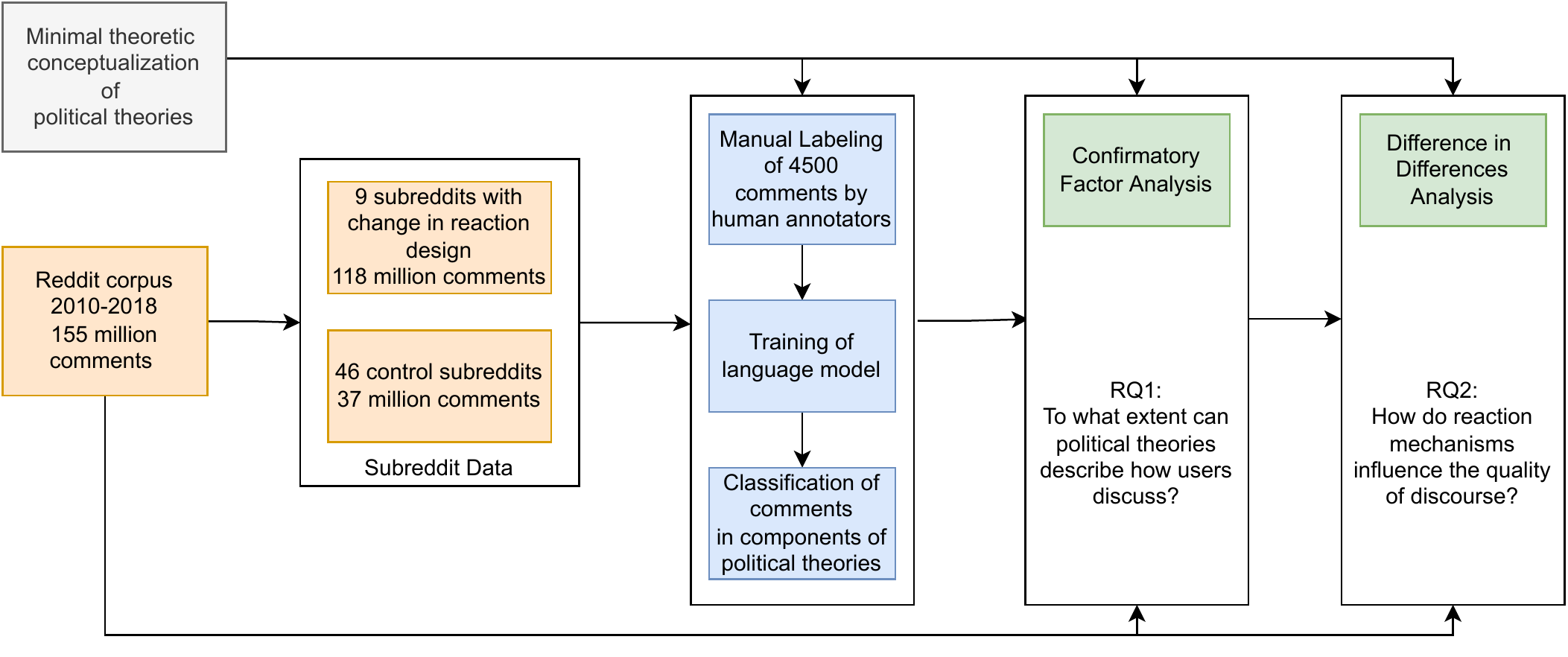}
    \caption{Overview of the study. We collect a sample of 155 million comments across 55 subreddits. Combining Confirmatory Factor Analysis and Difference in Differences Modeling, we investigate the extent to which political theories describe how users discuss political topics on Reddit (RQ1), and how the use of different reaction mechanisms correlates with an increase or decrease in the essential rhetoric elements of deliberative, civic, and demagogic discourse (RQ2).}
    \label{fig:overview}
\end{figure*}

Our study investigates to what extent the essential  components of prominent political discourse theories resurface in the political discussions of millions of users on Reddit. That is, do people's political conversations on Reddit incorporate the rhetoric components suggested by deliberative, civic, and demagogic discourse theory (\textbf{RQ1})?  We then test whether the existence of specific reaction mechanisms (i.e., upvotes, downvotes, no votes) correlates with deliberative, civic, and demagogic rhetoric components in political discussions on Reddit (\textbf{RQ2}). Combining an in-depth account on political discourse theories with a comprehensive, data-driven analysis of social media user discussions is necessary to best inform policy debates on the role of platform  reaction mechanisms in creating more civil, respectful, and just user discussion. Figure \ref{fig:overview} shows an overview of the entire study.

\section{Background \& related work}

\subsection{Understanding political discourse theories}

Political theorists and scientists develop analytic frames to understand how people speak when they discuss topics of political relevance  ~\cite{hicks2002promise, mccoy2002deliberative,sep-habermas,dahlgren2006civic}. In the last century, this line of scholarship has advanced prominent conceptions of political rhetoric, including the ones we study in this work (civic, deliberative, demagogic rhetoric). Deliberative discourse requires the giving and receiving of reasons when discussing propositions  ~\cite{sep-habermas}. In contrast, the rhetoric elements of civic discourse are less constrained by rationalization  ~\cite{barber1989public}. Demagogic speech tends to oversimplify complex societal issues  ~\cite{levinger2017love}. We provide an in-depth discussion of these three discourse theories in Section \ref{section_3}. 

\vspace{2mm}

Nonetheless, defining the exact demarcation lines between political discourse theories remains a contested terrain.  Political discourse theories are  ideal constructs and as such consist of constitutive  components that together \textit{ought} to represent  deliberative, civic, and demagogic discourse.   Through an in-depth engagement with scholarship on these three political discourse theories, we cast out their \textit{essential} rhetoric  components and explore to what extent these rhetoric  components can describe people's discussions of political topics on Reddit. We note that our analysis of these political discourse theories (and their subsequent application to social media user comments) necessarily falls back on our interpretation of the literature on political discourse theories. Clarifying our disciplinary backgrounds, our research team consists of political data scientists, philosophers, and computer scientists. While this allowed us to engage in recurring multidisciplinary discussion that mitigated possible biasing effects resulting from a discipline-specific reading of the literature, we wish to highlight that our interpretation of political discourse theories does not claim generalizability and, consequently, perfect external validity. Rather, our goal is to contribute to an ongoing policy discussion and we hope to encourage other scholars to replicate or otherwise perform similar research studies that attempt to bridge contested concepts of political discourse theories with empirical and quantitative analyses of social media discussions.

\subsection{Political discourse \& social media}

 Our study builds on and significantly extends research studies that have  performed first steps towards understanding social media users' political rhetoric. For example, using quantitative interviewing, Semaan et al.  ~\cite{semaan2014social} investigated whether social media users' interactions could be characterized by deliberative and civic agency. They described deliberation as the presence of reasoned and respectful discussions and civic agency as the ability to interact and participate in the public sphere.  Both Friess et al.  ~\cite{friess2021collective} and Wright et al.  ~\cite{doi:10.1177/1461444807081230} developed coding schemes for labeling user content as deliberative based on features such as rationality and constructiveness. Guimaraes et al.  ~\cite{guimaraes2019analyzing} formulated the conversational archetypes ``harmony'', ``discrepancy'', ``disruption'', and ``dispute'' to describe online political discourse. Lee et al.  ~\cite{lee2013does} connected user behavior on social media such as debating, posting or forwarding news, to features of civic engagement. Connecting online and offline behavior, Hampton et al.  ~\cite{hampton2017social} investigated the association of social media usage with the level of offline deliberation, which they defined as the propensity to discuss political issues with others. Evidently, prior research that investigates political rhetoric on social media has only used \textit{basic} conceptualizations of political discourse theories. We see this as an opportunity to perform a more in-depth engagement with scholarship on deliberative, civic, and demagogic discourse theories to understand the extent to which their essential  components map  to political discussions on Reddit.

\subsection{Reaction mechanisms, digital  environments, \& political discourse}
\label{affordances}

 Our study also  explores whether  specific digital  reaction mechanisms  on Reddit (i.e., upvoting and downvoting) relate to  the way users talk about political topics on social media. Previous studies have extensively analyzed how specific platform design features impact how users communicate with each other on the platform.  

\subsubsection{Platform design and content structure}
Focusing on design features that explicitly structure discussions, Rho et al.  ~\cite{10.1145/3313831.3376542} showed that the design input of political hashtags on social media influenced the deliberative quality of online discussions causing an increase in emotional and more black-and-white rhetoric. Kang et al. discuss various policy changes on the South Korean platform Naver including the removal of negative emoticons that aimed to reduce the amount of abusive and offensive comments ~\cite{kang2022closing}. Kriplean et al.  ~\cite{kriplean2012supporting} built a platform called ConsiderIt that helped users understand  previous user posts. By deploying list designs, the platform encouraged users to formulate pros and cons, leading to a higher level of deliberation. Furthermore, Aragón et al.  ~\cite{aragon2017thread} showed that  changing a linear to a hierarchical interface design increased social reciprocity on Menéame, a popular Spanish social news platform. In their study, \textcolor{black}{Wijenayake et al. ~\cite{wijenayake2020quantifying} manipulated user interactivity and response visibility in an online environment and found that these variables influence the level of conformity of users.} \textcolor{black}{Seering et al. found that presenting CAPTCHAs with positive stimuli to users leads them to externalize more positivity of tone and analytical complexity in their arguments ~\cite{seering2019designing}}. Liang et al.  ~\cite{liang2017knowledge} found that the maximum depth of a Reddit thread, and consequently of the respective discussions, was positively related to its rating (difference between up- to downvotes). In addition, Gilbert et al.  ~\cite{gilbert2013widespread} showed that users' tendency to focus on submissions that have higher rating on the platform resulted in an incidental ``filtering'' of information that would otherwise be of interest to them. 

\vspace{2mm}

\subsubsection{Reaction mechanisms and user behavior}
Besides design interventions that structure the content in a digital environment, many research studies have shown that reaction mechanisms have an impact on how users behave. Cheng et al.  ~\cite{cheng2014community} found that a higher number of downvotes across multiple social media platforms resulted in worsening the quality of discourse, while a higher count of positive reactions did not improve discourse significantly. Warut Khern-am-nuai et al.  ~\cite{khern2020haters} showed that after removing downvoting in a popular public forum, the number of both posts and replies significantly increased. Furthermore they found a decrease in toxicity and an increase in the diversity of replies. Shmargad et al.   ~\cite{shmargad2021social} demonstrated that upvoting incivility incentivized users to generate more toxic content. In a field experiment, Matias et al.  ~\cite{matias2018civilservant,Likeafox} showed that hiding downvotes increased the percentage of commenters who had not been vocal on political subreddits before. On Twitter, Adelani et al.  ~\cite{adelani2020estimating} demonstrated that user feedback expressed as likes and retweets significantly affected topic continuation in discussions. Stroud et al.  ~\cite{stroud2017like} concluded that the type of feedback that users gave by pressing a button (recommend, like, respect) altered the frequency and the scope of its use (see also  ~\cite{sumner2020assessing}). \textcolor{black}{Focusing on Reddit, Graham et al. ~\cite{graham2021sociomateriality} found that indeed users rarely use the voting reaction mechanisms as community guidelines dictate.} Generalizing, Hayes et al.  ~\cite{hayes2016one} found that users interpreted and applied the same reaction mechanism differently, depending on system, social, and structural factors.

Taken together, these studies underline that reaction mechanisms exert significant influence on user discourse in public online spheres. 
\subsubsection{Reaction mechanisms and recommender systems}
In addition to the direct impact of reaction mechanisms on social media discourse, reaction mechanism designs may have additional indirect consequences since user reactions are often used as signals in training data for recommendation algorithms, such as those used to order news feeds. Both existing  ~\cite{tiktok_2020} and proposed recommender systems  ~\cite{babaei2018purple,celis2019controlling,bountouridis2019siren} take different forms of user feedback as input to suggest content, be that likes, retweets, or other actions facilitated by reaction mechanisms. Such feedback does not always represent explicit user preferences about content, but rather is used as a way to overcome training issues of recommender systems. It is also useful for suggesting content that will keep users engaged, regardless of potential "spill over effects", i.e., users externalizing further unforeseen behaviors. ~\cite{zhao2018explicit,amatriain2009like,adomavicius2013recommender,hu2008collaborative}. 

\vspace{2mm}

Although user reactions can serve as a proxy, albeit imperfect, for user preferences where ground truth about these preferences is not available, the use of such proxies in training recommendation algorithms can also have undesirable effects. Prior research studies demonstrate that recommender algorithms' suggestions correlate with user radicalization ~\cite{ribeiro2020auditing},  discriminate against users and social groups  ~\cite{guo2021stereotyping}, and replicate political bias in discussions  ~\cite{papakyriakopoulos2020political,huszar2022algorithmic}. 
However, only few studies bridge between political discourse theories, design, and engineering  ~\cite{hampton2017social,lucherini2021t} to produce a better understanding of these phenomena.

\vspace{2mm}
\subsubsection{Reaction mechanisms as technical features}
\textcolor{black}{Before we engage in the theoretical discussion on the three political discourse theories, we need to point out the important conceptual distinction between \textit{reaction mechanisms} as technical features of a platform and platform \textit{affordances} as relational, community-specific behaviors that result from interacting with reaction mechanisms in a non-deterministic manner \cite{boyd2010social, treem2013social, evans2017explicating}. The technical features of digital artifacts (e.g., the downvoting functionality) are exactly the same to each user. However, different users may perceive and use such technical features differently resulting in the different interaction affordances that a common technical artifact provides to users. The key takeaway from this conceptual distinction is that the technical features of an artifact do not necessarily determine how users relate to and use the artifact. Thus, in our study, we refer to upvoting and downvoting as \textit{reaction mechanisms} because we do not explore how specific online communities and cultures differ in the perception, use, and interactions with such reaction mechanisms.}

\section{Minimal Conceptualization of political discourse theories} \label{section_3}

This study investigates essential rhetoric components of deliberation, civic engagement, and demagoguery. Foreshadowing stark differences, demagoguery oversimplifies complex societal issues  ~\cite{gustainis1990demagoguery, roberts2005democracy}. Demagoguery's rhetoric polarization lays the groundwork for \textit{action-based political mobilization}: you are either with ``us'' or with ``them''  ~\cite{hogan2006demagoguery,levinger2017love}. Civic engagement interactions are unstructured and characterized by multiple forms of communication and action, with its discourse being frequently characterized as ``messy conversation'' that facilitates participation  ~\cite{dahlgren2006civic}. Civic engagement interactions aim to be inclusive -- a key goal of civic engagement movements  ~\cite{mccoy2002deliberative}. Deliberation, from a Habermasian perspective, requires intersubjective propositional knowledge between conversation members  ~\cite{sep-habermas}. That is, knowledge-claims must fulfill standards of intersubjectivity: ideally, all discussants  are able to relate to the beliefs that underlie a proposition. Propositions must be grounded in logical plausibility, factual correctness, and communal narratives. Discussants' claims must have pragmatic value (i.e, ``propositionality'') in order to support the group's goal of reaching an understanding. Only when everyone can relate to the proposed statements, can deliberative rhetoric help solve a social, civil, or communal issue that is important to the group  ~\cite{hicks2002promise}. 

\vspace{2mm}

For our analysis of social media comments, we cannot measure whether a particular social media interaction leads to (or otherwise supports) specific offline actions. Our primary analysis seeks to investigate only the rhetoric components of these three political practices, that is, at the textual level of individual social media comments. We collect and analyze user comments from 55 political subreddits posted between 2010 and 2018. We presuppose a ``minimal conceptualization'' of demagogic, civic, and deliberative discourse. We here highlight this minimal conceptualization because we do not wish to suggest that an analysis of a textual corpus can fully account for the theoretical and practical complexities attributable to the theories. Nonetheless, it uncovers the extent to which the discursive elements of these theories manifest themselves in the political discussion on Reddit. 

\vspace{2mm}

In the following section, we outline the conceptual differences and commonalities of demagogic, civic, and deliberative discourse. This conceptual analysis serves to carve out the essential components of the three political discourse theories that we will use for data annotation and, eventually, classification of social media comments using the language model XLnet  ~\cite{yang2019xlnet}. 
\newline

Table \ref{labels} highlights the key rhetoric  components we develop for the labeling of social media comments and whether they are (indicated by a +) or are not (indicated by a -) part of demagoguery, civic engagement, or deliberation. In Appendix \ref{definitions_appendix}, we describe and justify the rhetoric  components in  more detail and provide further examples of social media comments. Finally, Appendix D, Table \ref{definitions} offers further explanations of the labels.

\subsection{Demagogic discourse} 

\textbf{Training set example comment}: \textit{``You socialist guys sound like superior human beings. I'm really impressed.''}

Demagogic discourse oversimplifies, distorts, or exaggerates complex societal challenges and has little regard for the truthfulness of propositions  ~\cite{gustainis1990demagoguery, roberts2005democracy}. In offering simple solutions that often entail ``pseudo-reasoning''  ~\cite{gustainis1990demagoguery}, demagogic statements are difficult to falsify if not even impervious and unresponsive to opposing arguments  ~\cite{roberts2005democracy}. Not only does this impede a constructive exchange of propositions but, for the demagogue, it renders perspective-taking of opposing positions irrelevant. The oversimplification of complex social phenomena results in a polarization that facilitates political mobilization  ~\cite{levinger2017love}. Demagogic talk aims to contrast social groups, highlighting apparent identity differences and putting them in competition with each other. It promises to care for the needs of ``ordinary'' people, creating an ethos around the often hateful division into laypeople and experts, elites and the forgotten, or poor and rich  ~\cite{roberts2005democracy, roberts2020demagoguery}. The disregard for truthfulness and the evocation of a collective identity based on hate, fear-mongering, and scapegoating means that demagogic rhetoric necessarily contains emotional linguistic  components  ~\cite{gustainis1990demagoguery}. For example, it typically expresses fear of outsiders and hatred against elites  ~\cite{roberts2020demagoguery, hogan2006demagoguery}.

\vspace{2mm}

Finally, demagogic rhetoric often speaks of a ``movement'' without specifying the particulars of its policy-making goals. As Levinger (2017) states, such movements rely on \textit{general} messages that tend to revolve around themes such as ``love for the country, its glorious past, degraded present, and utopian future.''  ~\cite{levinger2017love}. In contrast to civic and deliberative discourse, scholarship on demagogic rhetoric largely agrees on its constitutive  components. Its essential rhetoric devices are easier to pinpoint. It is important to note that demagogic political rhetoric and practice exist in many parts of the political spectrum and there are prominent cases for both left-leaning and right-leaning demagogues. Consequently, we do not argue (or implicitly suggest) that all demagogic rhetoric is necessarily right-wing or nationalist only. Demagoguery is an alienating discourse that appeals to the fancies and preconceptions of ``ordinary'' people, and it flourishes under different conditions and circumstances.  

\subsection{Civic discourse}
\label{civic}

\textbf{Training set example comment}: \textit{``Maybe we should look into why we're having more wildfires and address the issues that are causing that.''}

Civic engagement aims to mobilize people  to solve a commonly defined social or political issue. It allows for speech that remains unconstrained by ``overrationalization''  ~\cite{mccoy2002deliberative, barber1989public}. Several authors suggest that civic discourse considers rationality, neutrality, and a lack of emotional talk as \textit{hindrances} for multiple forms of speech  ~\cite{dahlgren2006civic, barber1989public}. ``Norms of deliberation'' and their associated speaking styles represent social privilege that can have a silencing effect for some participants  ~\cite{young2001activist}. However, the relationship between civic engagement and deliberation is not as clear cut. Some perspectives (for example,  ~\cite{adler2005we}) claim that civic engagement requires at least some deliberation to enable discussants to work towards a public goal. It needs to connect personal experience with public issues and thus encourages personal anecdotes, storytelling, or brainstorming  ~\cite{mccoy2002deliberative}. While civic engagement does not place priority on how participants formulate an argument and how well supported arguments are by evidence, it would be wrong to assume that the telling of personal experiences by participants does not contain any truthfulness. Indeed, if such personal anecdotes had no epistemic validity in the life of the community, then they could not produce a sense of connection and interrelatedness that is pivotal for civic engagement  ~\cite{diller2001citizens}.

\vspace{2mm}

Furthermore, civic discourse presupposes  a struggle or conflict for what a group considers to be a valuable civic goal. This requires discussants to listen to each other, take perspective, and critically analyze opposing arguments. As an inherently social practice, civic discourse represents a group's struggle to define political problems, draw up potential solutions, and mark out specific actions. This often involves intense exchange between engaged citizens that differ on and share perspectives on a single matter.  In civic engagement, interactions are both collaborative and competitive.

Successful public engagement processes evolve around a narrative of unity  ~\cite{hauser2004rhetorical}. In comparison to political parties and trade unions, mobilization  in civic engagement is oriented toward well-defined civil issues and causes  ~\cite{loader2014networked}. Online, the use of specific hashtagging, e.g., \#ferguson or \#policebrutality, creates a sense of a collective that often clearly demarcates ``who is with us'' and ``who is not''  ~\cite{bonilla2015ferguson}. Nonetheless, in civic discourse, a collective identity is built around a well-formulated cause. This stands in contrast to demagogic rhetoric that typically affirms a group's identity through the degradation of another group  ~\cite{levinger2017love}. 
Similarly, civic engagement addresses common concerns of a community. Participation aims at specific, practical, and do-able solutions  ~\cite{dahlgrenonline}, often externalized by protest movements calling for social action and change. This further differentiates civic from demagogic discourse, as the latter relies on more abstract and general messages promoted by social groups  ~\cite{levinger2017love}. Overall, civic discourse is an identity-based discourse whose success depends on a high level of empathy and practical, action-oriented incentives.

\subsection{Deliberative discourse}

\textbf{Training set example comment:} \textit{``I think those jobs should have a union, I'm in a white-collar union myself (as a teacher), and I have no idea why the private white collar sector shouldn't. I see corporations as an equal negotiation between management, shareholders, and employees, and all three should have a roughly equal stake, and that's only possible through unionization.''}

Different from demagoguery and civic engagement, deliberation rests on the ideal of reasoning, truth, and truthfulness  ~\cite{sep-habermas, hicks2002promise, young2001activist, halpern2013social}. Its rhetoric devices consist of logical reasoning and argumentation  ~\cite{halpern2013social}.  

While public reasoning does not (and cannot) fulfill the demands of scientific proof, ``(it) should not contradict the claims supported by the best available evidence''  ~\cite{hicks2002promise}: evidence that is publicly available and comprehensible for citizens. Besides drawing on the best available evidence, deliberative reasoning  requires interaction that presupposes  motivated participants that are able  to provide justifications for their assertions  ~\cite{young2001activist}. Deliberative discussions  aim to follow a particular structuring order. After rounds of debates, some members of the group may summarize others' claims and hence evaluate the considerations that speak in favor or against the presented propositions  ~\cite{halpern2013social}. 

\vspace{2mm}

Deliberative discourse works in the service of accomplishing a public goal that, eventually, should help improve participants' lives. Communicative practices  allow for and even encourage criticism of  other participants' arguments. However, counter argumentation is only legitimate when it rests on the premises and standards of public reasoning. Otherwise it ``trangresse(s) the limits of civility''  ~\cite{hicks2002promise}. In deliberative discourse, the ideal of reasoning is intimately connected to the moral principles of respect, equality, and trust  ~\cite{markovits2006trouble, sep-habermas}.  Such moral principles are often used to argue that deliberation is inclusive, a claim that has been met with scepticism by some authors  ~\cite{barber1989public, young2001activist}. 

\vspace{2mm}

Deliberative discourse, in contrast to demagogic and civic discourse, does not typically put emphasis on a collective identity. Rather, it presupposes that participants can move beyond their own interests and agree to work toward a common goal.  Participants eventually give up their own interest for the sake of a collective identity in civic discourse. In contrast, in deliberative discourse, rhetoric demands for reasoning and truthfulness are supposed to put strong normative pressure on discussants’ interactions. Hicks (2002) states that ``citizens agree to justify their political proposals...because they agree to propose and abide by the terms of fair cooperation...they will accept the results of public deliberation as binding and agree to abide by those results even at the costs of their own interests.''  ~\cite{hicks2002promise}. Consequently, deliberative discourse is strongly based on the factuality of content, argumentative completeness, and respect of other discussants.

\subsection{Mapping theories to essential components}
\label{mapping}

 Our previous discussion demonstrates some of the conceptual plurality inherent to different political discourse theories. After an in-depth engagement with and critical reading of the literature, and several subsequent rounds of discussion among co-authors, we argue that there is sufficient agreement among scholars on the essential components of each type of political discourse to train a classifier that is able to discriminate among them. Developing the corresponding set of labels was a cyclical rather than a linear process. After critical engagement with the cited literature on the three political discourse theories, two co-authors separately developed codes based on their analysis of the most essential components of each political discourse theory. Then, they compared the created categories and, with the aid of a third co-author, agreed on an initial set of components.  Through multiple rounds of discussion, two co-authors assigned the essential components to each of the three discourse theories. This process led to further discussion on the \textit{definitional scope} of the components. Thus, through multiple rounds of discussion between three co-authors, going back and forth between the choices of essential components and their assignment to the three discourse theories, we finally agreed on thirteen essential components that could be operationalized in a multilabel classification task (see final set of components together with their definitions in Appendix \ref{definitions_appendix} \& Table \ref{definitions}).  We document disagreement among co-authors on the definition and assignment of some of the components in Appendix \ref{definitions_appendix} (see Fact-related argument in Appendix \ref{fact_related} and Identity Labels in Appendix \ref{identity}). Table \ref{labels} presents an overview of the essential components and how we assigned them to each political discourse theory for our multilabel classification task.

\begin{table}
\caption{Rhetoric  components of the minimal theoretic conceptualization of deliberative, civic, and demagogic discourse.}
\label{labels}

\resizebox{1\columnwidth}{!}{\begin{tabular}{|p{5cm}|p{5cm}|}
 \hline
 \multicolumn{2}{|c|}{\textbf{Deliberative discourse}} \\
 \hline
 Argument is part of theory (+) & Argument is not part of theory (-) \\
 \hline
 fact-related argument  ~\cite{hicks2002promise, markovits2006trouble, gastil2000face, sep-habermas, black2011self} & we vs. them  ~\cite{sep-habermas}   \\
 structured argument  ~\cite{hicks2002promise, markovits2006trouble, black2011self, robertson2010off}  & generalized call for action  ~\cite{yang2019xlnet, hicks2002promise, markovits2006trouble}   \\
 counterargument  ~\cite{sep-habermas, halpern2013social}  & who instead of what  ~\cite{sep-habermas, hicks2002promise}  \\
 empathy/reciprocity  ~\cite{sep-habermas, young2001activist}    & emotional language  ~\cite{hicks2002promise, halpern2013social, sep-habermas}  \\
  & unsupported argument  ~\cite{hicks2002promise, halpern2013social, sep-habermas}     \\
  &   \\
 \hline
\end{tabular}}
\vspace{3mm}

\resizebox{1\columnwidth}{!}{\begin{tabular}{|p{5cm}|p{5cm}|}
 \hline
 \multicolumn{2}{|c|}{\textbf{Civic discourse}} \\
 \hline
 Argument is part of theory (+) & Argument is not part of theory (-) \\
 \hline
 situational call for action  ~\cite{adler2005we, dahlgren2006civic, bonilla2015ferguson, skoric2016social} & fact-related argument  ~\cite{barber1989public, mccoy2002deliberative, dahlgren2006civic}   \\
 we vs. them  ~\cite{loader2014networked, mccoy2002deliberative, dahlgren2006civic, bonilla2015ferguson}  & structured argument  ~\cite{barber1989public, mccoy2002deliberative, dahlgren2006civic}    \\
 counterargument  ~\cite{barber1989public, mccoy2002deliberative} & generalized call for action  ~\cite{adler2005we, dahlgren2006civic}  \\
 empathy/reciprocity  ~\cite{young2001activist, dahlgren2006civic, mccoy2002deliberative, barber1989public}    &  \\
 emotional language  ~\cite{young2001activist, dahlgren2006civic, mccoy2002deliberative, barber1989public}   &     \\
 collective rhetoric  ~\cite{mccoy2002deliberative, dahlgren2006civic, bonilla2015ferguson} &    \\
   &   \\
 \hline
\end{tabular}}

\vspace{3mm}

\resizebox{1\columnwidth}{!}{\begin{tabular}{|p{5cm}|p{5cm}|}
 \hline
 \multicolumn{2}{|c|}{\textbf{Demagogic discourse}} \\
 \hline
 Argument is part of theory (+) & Argument is not part of theory (-) \\
 \hline
 you in the epicenter  ~\cite{levinger2017love, gustainis1990demagoguery}  & fact-related argument  ~\cite{gustainis1990demagoguery, roberts2005democracy, roberts2020demagoguery}   \\
 we vs. them  ~\cite{hogan2006demagoguery, roberts2020demagoguery} & structured argument  ~\cite{gustainis1990demagoguery, roberts2005democracy, roberts2020demagoguery}    \\
 generalized call for action  ~\cite{roberts2020demagoguery, hogan2006demagoguery} & empathy/reciprocity  ~\cite{levinger2017love, gustainis1990demagoguery}  \\
 who instead of what  ~\cite{gustainis1990demagoguery}    & counterargument \cite{roberts2020demagoguery, roberts2005democracy}  \\
 emotional language  ~\cite{roberts2005democracy, gustainis1990demagoguery} &     \\
 unsupported argument  ~\cite{roberts2005democracy, gustainis1990demagoguery} &    \\
 collective rhetoric  ~\cite{roberts2020demagoguery, hogan2006demagoguery, gustainis1990demagoguery}  &   \\
 \hline
\end{tabular}}
\end{table}

\subsection*{Essential components of political discourse}

First, \textit{demagogic discourse} is represented by the presence of collective, emotional and ``we vs. them'' rhetoric, unsupported arguments (including those who prime identity and group membership), and calls for generalized abstract action. It does not include fact-related, structured and empathetic arguments. ``Collective rhetoric'' classified comments that emphasize a collective identity around words such as ``we'' or ``our'' in a way that promotes group membership. 

\vspace{2mm}

Second, in \textit{civic discourse}, the essential components include emotional, collective, and ``we vs. them'' rhetoric, together with counterarguments, and statements that call for situational action. In contrast, civic discourse explicitly excludes fact-related, structured arguments, and calls for generalized abstract action.
The label ``we vs. them'' classified user comments that contrasted, discriminated or degraded another group with the purpose to consolidate the identity of the user's group. In contrast to the concept ``generalized call for action'', ``situational call for action'' included comments that described a specific policy goal. 

\vspace{2mm}

Third, \textit{deliberative discourse} includes fact-related and structured arguments, counterargument statements, the presence of empathy/reciprocity in user comments, as well as the explicit absence of rhetoric that does not provide any evidence (unsupported statements or statements that focus on who is doing something instead of what is happening, emotional language, and generalized abstract calls for action). The concept ``fact-\textit{related} argument'' consisted of two types of justifications: empirical and reasoned justifications. When supporting a claim, empirical justifications provided either a direct reference to other sources (e.g., in the form of links or article references) or referred to personal experiences and anecdotes relevant for the overall claim. With the label ``empathy/reciprocity'' we classified comments that explicitly acknowledged another user's perspective, claim, or proposition. The identity label ``who instead of what'' classified comments that referred to a person or social group without specifying any of their behavior or action. In comments with the label ``generalized call for action'' users stated an explicit need for policy change without providing any justification. 

\vspace{2mm}

A more detailed description and justification of \textit{all} argument types, together with examples and explanations for the generation of the minimal conceptualization can be found in Appendix \ref{definitions_appendix}. Next, we use these essential components as labels for the classification of comments on Reddit to quantify the prevalence of the political discourse theories on the platform.

\section{Data \& Methods}

\subsection{Data Collection}
To understand whether and to what extent essential components of political discourse theories  characterize political discussions on social media, we collected a large volume of user comments from Reddit. We decided to conduct our study on Reddit for three reasons. First, Reddit offers free full access to historical data of the platform. Hence, we were able to perform a large-scale data analysis at the level of an entire ecosystem  ~\cite{zuckerman2021study}. Second, Reddit discussions are hosted in separate message boards (i.e., subreddits), with each of them having a specific topic of discussion. This allowed us to focus on subreddits with discussions of political topics. Third, Reddit enables moderators of each subreddit to customize their subreddit's user interface, \textit{allowing them to select which reaction mechanisms (upvote, downvote, or both)  are available to  users}. Therefore, Reddit provided an ideal ecosystem to perform our analysis. 

\vspace{2mm}

To create our pool of Reddit comments, we first generated a list of 55 political subreddits, nine of which changed their available reaction mechanisms at some point (retrieved from  ~\cite{reddit_reaction}). Then, we used the pushshift API  ~\cite{baumgartner2020pushshift} to extract all comments created in these subreddits between January 1, 2010 and April 1, 2018. We chose this time frame since many of the subreddits in our sample were created around 2010. Furthermore, moderators have the ability to customize the old Reddit interface, which stopped being the default interface on April 2, 2018  ~\cite{liao_2018}. Overall, we collected 155 million comments created during this period. We used the Wayback Machine  ~\cite{internetarchive} to extract the intervals at which each subreddit introduced a change in their available reaction mechanisms. Table \ref{subreddits} in Appendix \ref{tables_appendix} offers an overview of the subreddits used in the study, and Table \ref{reactions} in Appendix \ref{tables_appendix} indicates the date when nine of the political subreddits changed their available reaction mechanisms. 


\begin{table}
\caption{The number of times coders assigned each label to comments in the sample together with the language model's F1 score for each class. }
\label{comments_f1}
\resizebox{1\columnwidth}{!}{
\begin{tabular}{rcccc}
\textbf{Label}              & \begin{tabular}[c]{@{}c@{}}  \textbf{N. of}\\  \textbf{occurences}  \end{tabular} & & \begin{tabular}[c]{@{}c@{}} \textbf{Recall} \\ \textbf{evaluation} \end{tabular} &  \begin{tabular}[c]{@{}c@{}} \textbf{F1 score} \\ \textbf{evaluation} \end{tabular} \\ \hline
You in the epicenter        & 300.0                     &  &   \textcolor{black}{0.82}  &   \textcolor{black}{0.86}            \\
We vs. them                    & 366.0                     &  &   \textcolor{black}{0.68} &  \textcolor{black}{0.79}             \\
Generalised call for action & 371.0                     &  &   \textcolor{black}{0.87} & \textcolor{black}{0.90}             \\
Situational call for action & 304.0                     & &   \textcolor{black}{0.96}  & \textcolor{black}{0.96}             \\
Who instead of what         & 345.0                     &  &   \textcolor{black}{0.87}   &  \textcolor{black}{0.90}           \\
Fact-related argument       & 1307.0                    &  &   \textcolor{black}{0.89}  & \textcolor{black}{0.78}             \\
Structured argument         & 1124.0                    &  &   \textcolor{black}{0.90}  & \textcolor{black}{0.88}             \\
Counter-argument structure  & 563.0                     &  &   \textcolor{black}{0.91}  &  \textcolor{black}{0.83}            \\
Empathy/reciprocity         & 329.0                     &  &   \textcolor{black}{0.90}  & \textcolor{black}{0.86}             \\
Emotional language          & 438.0                     &  &   \textcolor{black}{0.74}  & \textcolor{black}{0.77}             \\
Collective rhetoric         & 469.0                     &  &   \textcolor{black}{0.81}  & \textcolor{black}{0.84}             \\
Unsupported argument        & 422.0                     &  &   \textcolor{black}{0.67}  &  \textcolor{black}{0.76}             \\
Other                       & 874.0                     &  &   \textcolor{black}{0.91}  &  \textcolor{black}{0.84}           \\ \hline
\end{tabular}}
\end{table}

\subsection{Annotation \& model training}
To map our minimal conceptualization of deliberative, demagogic, and civic discourse to discussions on Reddit, we labeled a set of comments from our Reddit corpus and trained a large language model in a multilabel classification task. Two coders labelled 4,500 unique comments with at least one of the thirteen types of essential components extracted from our theoretical analysis (see label development documentation in Section \ref{mapping} and Appendix \ref{definitions_appendix}). To ensure intercoder reliability \textcolor{black}{and the representativeness of the sample}, we performed the following procedures. First, coders discussed the developed minimal theoretic conceptualization, reviewed predefined examples for each class, and resolved any questions about the nature of the classes. Then, both coders labelled the same set of \textcolor{black}{100 random} comments from the corpus, yielding a Krippendorf alpha of 0.7. After discussing prevailing differences in the coding tactics, coders learned to adjust their coding in a way that conformed more to the theoretic framework.  Coders \textcolor{black}{ then} relabelled the same corpus, yielding an intercoder reliability of 0.75. This was higher than the expected minimum for coding complicated language tasks in the literature (>0.6  - see e.g.,  ~\cite{sap2019social,morstatter2018identifying,daxenberger2012corpus,amidei2019agreement, engelmann2022people, ullstein2022ai}). Since coders' labeling practices were robust, they proceeded with the  \textcolor{black}{selection and labeling of further comments. Specifically, an initial sample of 1650 comments was labeled, containing 30 comments from each subreddit in the dataset. Next, a second batch of 1000 comments was annotated, which we randomly selected by stratified sampling among subreddits. To assess how many comments were necessary for an accurate classifier, we trained a preliminary language model on these 2750 comments, finding that for reliable prediction for each class, at least 300 observations were necessary.  To satisfy this condition, we continued labeling comments by stratified random sampling until an annotated sample of 4500 comments was produced.}

\vspace{2mm}

\textcolor{black}{For the final model,} we split the corpus of the 4500 comments in a 80-20 train/test set. We kept capitalization and punctuation of comments in their original form and removed the quoted content in the case that a user was quoting another user. To train our model, we used the large language model \textit{XLnet}  ~\cite{yang2019xlnet}. XLnet is an architecture that combines transformers and auto-regressive modeling. \textcolor{black}{We selected XLnet over other commonly used language models such as BERT  ~\cite{devlin2018bert} because it still holds the top performance in multiple text classification benchmarks (e.g., first place in Amazon-5, Amazon-2, DBpedia, Yelp-2, AG News, second place in IMDb, Yelp-5)	 ~\cite{paperswithcodePapersWith}. } We applied a warm-up initialization of 0.1, a learning rate of 3e-5, and a maximum sequence length of 100 words. 
Our final model resulted in a label ranking average precision score of 91\%, while all class specific F1 scores were higher than 0.87. \textcolor{black}{To ensure the robustness of our model, we additionally created an evaluation set of 200 comments, in which each class of the dataset appeared at least ten times. On the evaluation set, the model achieved an accuracy of 0.72, label ranking average precision score of 0.85, while all class specific F-1 scores were higher than 0.76 (Figure \ref{comments_f1}).  } Given the obtained model accuracy, we then  analyzed a total of 155 million comments in our corpus.

\section{Answering RQ 1: To what extent can essential rhetoric components of deliberative, civic, and demagogic discourse describe users' political discussions on Reddit?}


\subsection{Confirmatory factor analysis}

To understand to what extent the essential components of deliberation, civic discourse, and demagoguery  characterize social media discussions, we conducted a confirmatory factor analysis (CFA). \textcolor{black}{CFA is generally used to test hypotheses about plausible model structures  ~\cite{bollen1989structural}, and has commonly been used to model different types of data, from survey information to time-series  ~\cite{bollen2006latent}. Until now, the application of CFA in NLP-driven questions has been limited (e.g.,  ~\cite{park2017rotated}), and our study serves as an inspiration for exploiting its capabilities, but also understanding its limits in machine-learning based research.} Factor analysis allowed us to mathematically represent the political theories as latent unobserved variables (factors) as described by a set of observed variables (items). The observed variables corresponded to the different rhetoric components as predicted by the language model. We chose CFA rather than exploratory factor analysis (EFA) since we were investigating whether a specific  conceptualization of discourse theories  empirically characterized social media discussions, and not which general argument structures were best described by our data. With CFA, we could construct structures of variables that complied with the minimal conceptualizations of political theories and by assessing the quality of model-fit, we explored to what extent political theories can describe how users  discuss political topics on Reddit. Furthermore, we quantified which arguments were empirically associated with which theories and their corresponding magnitude of importance. 

\vspace{2mm}
\begin{algorithm*}
\caption{CFA model selection}
\label{algorithm}
\begin{algorithmic}[1]
\Procedure{select\_model}{}
\State $ best_{model}  \leftarrow NULL$  
\State $ best_{CFI}  \leftarrow  0 $ 
\State $ expansive_{model}  \leftarrow NULL $ 
\State $ expansive_{CFI}  \leftarrow   0 $
\State $ $
\color{black}
\State $ discourses  \leftarrow   (deliberative,civic,demagogic) $
\State $ $
\State $ theory(demagogic)  \leftarrow   (you\ in\ the\ epicenter,we\ vs\ them,generalized\ call,who\ instead\ of\ what,$
\State $emotional\ language,\ unsupported\ argument,\ collective\ rhetoric,\ fact\ related\ argument, structured\ argument,$ 
\State $empathy/reciprocity,\ counterargument) $
\State $ $
\State $ theory(civic)  \leftarrow   (situational\ call,we\ vs\ them,\ counterargument,\ empathy/reciprocity,$
\State $emotional\ language,\ collective\ rhetoric,\ fact\ related\ argument, structured\ argument,\ generalized\ call) $
\State $ $
\State $ theory(deliberative)  \leftarrow   (fact\ related\ argument,\ structured\ argument,\ counterargument,$
\State $empathy\ reciprosity,\ we\ vs\ them,\ generalized\ call\ ,who\ instead\ what, emotional\ language,\ unsupported\ argument) $
\State $ $
\color{black}
\For{\texttt{discourse} in \texttt{discourses}}
    \State \texttt{$loadings_{discourse}  \leftarrow 0$}
\EndFor
\State $ $
      \For{\texttt{argument} in \texttt{arguments}}
            \For{\texttt{discourse} in \texttt{discourses}}
    \If{\texttt{argument} in \texttt{theory(discourse)}}
\State $ $
    \State $ loadings_{discourse} += argument $
    
    \State $ model \leftarrow  calculate\_cfa(loadings)$
\State $CFI, errors \leftarrow get\_model\_metrics(model) $
      \If{$errors=0$}
  \State $expansive_{model}  \leftarrow model $
    \State$ expansive_{CFI}  \leftarrow CFI $
      \EndIf
\If{$CFI > best_{CFI}$ and $errors=0$ }
  \State $best_{model}  \leftarrow model $
    \State$ best_{CFI}  \leftarrow CFI $      
    \EndIf    
    \EndIf 
    \EndFor 
    \EndFor 
      \EndProcedure
\end{algorithmic}
\end{algorithm*}
We created CFA models in which each political theory was represented as a function of all or a subset of the arguments that compose it, following the minimal conceptualization we performed (see Table 1). We did so by applying the algorithm depicted in Algorithm \ref{algorithm}. We initiated a null model in which the latent factors are not loaded by any variable, and by iterating over the list of argumentations, we added them to the equation of theories that are associated with them. \textcolor{black}{An argumentation was associated with a specific theory, if according to the theory it explicitly appears in it or is absent from it. For example, both emotional rhetoric and structured argument are associated with demagogic discourse, because the theory dictates that emotional language is its constitutive element, but also well-formed arguments are absent from it. Therefore, we hypothesized a positive loading of the emotional rhetoric item and a negative loading of the structured argument item on the factor of demagogic rhetoric. In contrast, counterargument structure is not associated with demagogic rhetoric according to the theories we engaged with  and hence we did not load it on the factor at all.} For each addition of an argumentation, we calculated the CFA model and stored its Confirmatory Fit Index (CFI) score  ~\cite{bentler1980significance} as a metric for model fit. After finishing the process, we selected two models. We kept the model with the highest CFI, which converged without errors, and the one with the most loaded arguments that converged without errors. The first model revealed those elements of the theories that were used in discussions on Reddit, while the second described how good the closest empirical model to the theories described political discussions on Reddit. 

\vspace{2mm}

For all models we created, we allowed factors to correlate since discourse theories shared commonalities and differences in their  essential components. Thus, we also included cross-loadings in our models. Moreover, we did not use a cut-off when evaluating the magnitude of factor loadings. Studies suggest to use a cut-off given a small sample size  ~\cite{guadagnoli1988relation,stevens2012applied} (e.g., less than 1000 observations) to avoid type I or type II errors. In our case, the sample size was at the magnitude of millions, yielding all detected associations statistically significant. Therefore, our criterion for evaluating a factor loading was theory-driven only. Each discourse theory was described by at least nine elementary arguments, and a user comment would rarely include more than two argumentation types at the same time, as most comments on Reddit do not exceed two to three sentences. Thus, \textcolor{black}{besides dominant loadings with high coefficients,  we did not automatically reject loadings with low values (even <0.2), as even arguments appearing sparsely could be plausible and theory-conforming.} A simulation that showed this case can be found in the Appendix \ref{simulation}. \textcolor{black}{ In contrast, our model selection process was primarily informed by overall model fit, as adding variables that did not comply to the model structure would result in lower values of CFI and TLI.}

Next, after selecting the best and most expansive model, and by drawing from the developed theoretic framework, we sought to answer whether political discourse theories could describe how people talked on political subreddits between 2010-2018.

\subsection{Results for RQ1}

The CFA results show that the discourse theories  characterize political discussions on Reddit, aligning to a large extent with the conceptualizations of political theorists. Deliberative, civic, and demagogic discourse were present in the subreddits we studied, with user discourse related comments containing rhetoric  components belonging to the theories. Table \ref{loadings} presents which items (arguments) loaded to each theory for the best and most expansive model \textcolor{black}{ (see also figures with diagrams \ref{fig:robust},\ref{fig:expansive} in the Appendix )}, while Table \ref{model_fit} provides the corresponding Goodness of Fit metrics. 

\begin{table}
\caption{Fit of the best and most expansive CFA models assessing the prevalence of demagogic, civic, and deliberative discourse on our Reddit sample.}
\label{model_fit}
\resizebox{0.45\textwidth}{!}{\begin{tabular}{c|cc}
      & Best model & Expansive model \\ \hline
\begin{tabular}[c]{@{}c@{}} Confirmatory Fit Index\\ (CFI)  \end{tabular} & 0.97       & 0.85            \\[0.3cm]
\begin{tabular}[c]{@{}c@{}}Tucker Lewis Index \\ (TLI) \end{tabular}  & 0.95      & 0.74            \\[0.3cm]
\begin{tabular}[c]{@{}c@{}}Root Mean Square Error \\ of Approximation (RMSEA) \end{tabular} & 0.027      & 0.044           \\[0.3cm]
\begin{tabular}[c]{@{}c@{}}Standardized Root Mean Square \\Residual (SRMR)\end{tabular}  & 0.016      & 0.028          \\[0.3cm] \hline
\end{tabular}}
\end{table}

\vspace{2mm}

The best model showed a very good fit (CFI$>$0.95) and contained at least three items for each discourse construct. When users performed deliberative discussions, they created fact-related and structured arguments, with the variables' loadings being 0.743 \& 0.691, respectively. Furthermore, they avoided generating unsupported statements ($\beta = -0.202$). Civic discourse on Reddit was  characterized mostly through statements of collective rhetoric ($\beta = 0.891$) together with statements calling for situational action  ($\beta = 0.194$) and ``we vs. them'' statements   ($\beta = 0.211$). 

\vspace{2mm}

As expected, ``we vs. them'' rhetoric was also present in users'  demagogic discussions ($\beta = 0.165$), albeit the strongest argumentation type in it was the generation of emotional comments ($\beta = 0.485$), followed by unsupported arguments  ($\beta = 0.235$) and ``who instead of what'' rhetoric  ($\beta = 0.165$).  Factors covaried pairwise (demagogic \& civic $\beta =-0.555$, civic \& deliberative  $\beta =0.418$, demagogic \& deliberative $\beta =-0.054$), as discourses shared specific argumentation types.  

\vspace{2mm}

Focusing on the expansive model, it included 24 out of the 28 theoretical item-factor associations. Again, factors correlated pairwise (demagogic \& civic $\beta =-0.213$, civic \& deliberative  $\beta =0.528$, demagogic \& deliberative $\beta =-0.214$). Nonetheless, the model's fit was borderline, since RMSA was acceptable (0.044 < lower than 0.05), but CFI was not (0.85, lower than the generally recommended threshold of 0.9)  ~\cite{xia2019rmsea}. For the additional variables that were not included in the best model, they generally had weak associations with the constructs, except of four cases. For civic discourse, ``empathy/reciprocity'' was associated to the construct ($\beta = 0.234$), while collective rhetoric loaded on demagogic discourse ($\beta = 0.227$). These associations complied with theoretical conceptions of the theories. In contrast, empathy/reciprocity ($\beta = -0.173$) and generalised call for action ($\beta = -0.132$) loaded on deliberative and demagogic discourse, respectively. This contradicted the theoretical conceptualization of these two discourse theories. 

\begin{table*}
\caption{Magnitude of factor loadings for the expansive and best CFA model. For each discourse type (deliberative, civic, demagogic) we provide which factors are  (+) or are not (-) constitutive  components. If a factor is part of a model and complies with the minimal theoretic conceptualization it is colored in \setlength{\fboxsep}{0.8pt}\colorbox{green!25}{green}. If it is part of the a model but contradicts the theororetic conceptualization it is colored in \setlength{\fboxsep}{0.8pt}\colorbox{red!25}{red}. If we were not able to include it in a error-free model it is colored in \setlength{\fboxsep}{0.8pt}\colorbox{orange!50}{orange}.  }
\label{loadings}
\begin{flushleft}

\resizebox{0.93\textwidth}{!}{\begin{tabular}{r@{\hskip1mm}|c@{\hskip2mm}c@{\hskip2mm}c@{\hskip2mm}c@{\hskip2mm}c@{\hskip2mm}c@{\hskip2mm}c@{\hskip2mm}c@{\hskip2mm}c@{\hskip2mm}}
   \begin{tabular}[r]{@{}r@{}}\textbf{Deliberative}\\\textbf{ discourse}\end{tabular}              & \multicolumn{9}{c}{}                                                                                                                                                                                 \\ \hline \hline
Argumentation   & \begin{tabular}[c]{@{}c@{}}fact-related\\argument \end{tabular} &\begin{tabular}[c]{@{}c@{}} structured\\argument\end{tabular} & \begin{tabular}[c]{@{}c@{}}counter\\argument \end{tabular}&\begin{tabular}[c]{@{}c@{}}empathy\\reciprocity \end{tabular} & \begin{tabular}[c]{@{}c@{}}we vs\\them \end{tabular}& \begin{tabular}[c]{@{}c@{}}generalised\\call\end{tabular} & \begin{tabular}[c]{@{}c@{}}who instead\\of what \end{tabular}&\begin{tabular}[c]{@{}c@{}} emotional\\language \end{tabular}& \begin{tabular}[c]{@{}c@{}}unsupported\\argument \end{tabular}\\ \hline
Theory          & +                    & +          & +                & +                   & -                  & -                    & -                   & -                     & -                                   \\
Expansive model & \cellcolor{green!25}0.677              &\cellcolor{green!25} 0.7772     & \cellcolor{green!25}0.005           &
\cellcolor{red!25}-0.173      & \cellcolor{orange!50}              & 
\cellcolor{red!25}0.082               & 
\cellcolor{green!25}-0.072               & 
\cellcolor{green!25}-0.083              &  \cellcolor{green!25} - 0.225               \\
Best Model      &  \cellcolor{green!25} 0.743 & \cellcolor{green!25} 0.691         &    \cellcolor{orange!50}              &     \cellcolor{orange!50}                &          \cellcolor{orange!50}          &  \cellcolor{orange!50}                    &          \cellcolor{orange!50}           &            \cellcolor{orange!50}           &    \cellcolor{green!25} -0.202      \\ \hline                              
\end{tabular}}
\newline
\vspace*{0.3 cm}
\newline
\resizebox{0.93\textwidth}{!}{\begin{tabular}{r@{\hskip1mm}|c@{\hskip3mm}c@{\hskip3mm}c@{\hskip3mm}c@{\hskip3mm}c@{\hskip3mm}c@{\hskip3mm}c@{\hskip3mm}c@{\hskip3mm}c@{\hskip3mm}}
         \begin{tabular}[r]{@{}r@{}}\textbf{Civic}\\\textbf{ discourse}\end{tabular}       & \multicolumn{9}{l}{}                                                                                                                                                                                 \\ \hline \hline
Argumentation   & \begin{tabular}[c]{@{}c@{}}situational\\call \end{tabular} &\begin{tabular}[c]{@{}c@{}} we vs. \\them\end{tabular} & \begin{tabular}[c]{@{}c@{}}counter\\argument \end{tabular}&\begin{tabular}[c]{@{}c@{}}empathy\\reciprocity \end{tabular} & \begin{tabular}[c]{@{}c@{}}emotional\\language \end{tabular}& \begin{tabular}[c]{@{}c@{}}collective\\rhetoric \end{tabular}& \begin{tabular}[c]{@{}c@{}}fact-related\\argument\end{tabular} & \begin{tabular}[c]{@{}c@{}}structured\\argument \end{tabular}& \begin{tabular}[c]{@{}c@{}}generalized\\call \end{tabular}\\ \hline
Theory          & +                    & +          & +                & +                   & +                 & +                    & -                   & -                     & -                                    \\
Expansive model & \cellcolor{green!25}0.162               &\cellcolor{green!25} 0.208    & \cellcolor{orange!50}           &
\cellcolor{green!25}0.2347      & \cellcolor{green!25}  0.062            & 
\cellcolor{green!25}0.494               & 
\cellcolor{orange!50}              & 
\cellcolor{green!25}-0.037              &  \cellcolor{green!25} - 0.098               \\
Best Model      &       \cellcolor{green!25} 0.194             &    \cellcolor{green!25}  0.211       &      \cellcolor{orange!50}             &        \cellcolor{orange!50}           &             \cellcolor{orange!50}        &                \cellcolor{green!25}  0.891     &                 \cellcolor{orange!50}     &                    \cellcolor{orange!50}    &                                  \cellcolor{orange!50}         \\ \hline      
\end{tabular}}

\vspace*{0.3 cm}

\resizebox{0.99\textwidth}{!}{\begin{tabular}{r@{\hskip3mm}|c@{\hskip2mm}c@{\hskip2mm}c@{\hskip2mm}c@{\hskip2mm}c@{\hskip2mm}c@{\hskip2mm}c@{\hskip2mm}c@{\hskip2mm}c@{\hskip2mm}c@{\hskip1mm}c@{\hskip1mm}}
       \begin{tabular}[r]{@{}r@{}}\textbf{Demagogic}\\\textbf{ discourse}\end{tabular}         & \multicolumn{11}{c}{}                                                                                                                                                                                 \\ \hline \hline
Argumentation   & \begin{tabular}[c]{@{}c@{}}you in the\\ epicenter \end{tabular} &\begin{tabular}[c]{@{}c@{}} we vs\\ them\end{tabular} & \begin{tabular}[c]{@{}c@{}}generalised\\ call \end{tabular}&\begin{tabular}[c]{@{}c@{}} who instead\\ of what \end{tabular} & \begin{tabular}[c]{@{}c@{}}emotional\\ language \end{tabular}& \begin{tabular}[c]{@{}c@{}}unsupported\\ argument\end{tabular} & \begin{tabular}[c]{@{}c@{}}collective\\ rhetoric \end{tabular}&\begin{tabular}[c]{@{}c@{}} fact-related\\ argument \end{tabular}& \begin{tabular}[c]{@{}c@{}}structured\\ argument \end{tabular}&\begin{tabular}[c]{@{}c@{}} empathy\\reciprocity \end{tabular}&\begin{tabular}[c]{@{}c@{}} counter\\argument \end{tabular}\\ \hline
Theory          & +                    & +          & +                & +                   & +                  & +                    & +                   & -                     & -                   & -                   \\
Expansive model & \cellcolor{red!25}-0.025               &\cellcolor{green!25} 0.223      & \cellcolor{red!25}-0.132            & \cellcolor{green!25}0.142               & \cellcolor{green!25}0.453              & \cellcolor{green!25}0.291                & \cellcolor{green!25}0.227               & \cellcolor{green!25}-0.052                 &    \cellcolor{orange!50}                 &\cellcolor{green!25} -0.063 & \cellcolor{orange!50}                   \\
Best Model      &    \cellcolor{orange!50}                 &     \cellcolor{green!25} 0.133       &        \cellcolor{orange!50}           &     \cellcolor{green!25} 0.165                &   \cellcolor{green!25}   0.485               &   \cellcolor{green!25}   0.235                &    \cellcolor{orange!50}              &    \cellcolor{orange!50}                    &    \cellcolor{orange!50}                  &      \cellcolor{orange!50}      & \cellcolor{orange!50}  \\ \hline                
\end{tabular}}
\end{flushleft}

\end{table*}
\vspace{2mm}

The CFA's results reveal that there is a clear connection between theory  and social media discussions, and also show which key rhetoric components of deliberative, civic, and demagogic discourse describe user interactions in our sample (RQ1). 
Although user comments included central features of each discourse, such as fact-related and structured arguments in deliberation, collective rhetoric in civic discourse, and unsupported, emotional, and identity-related (``we vs. them'') statements for demagogic discourse, other properties prescribed by theorists to each discourse did not empirically connect to the latent constructs. Nonetheless, there is a sufficient overlap between theoretical conceptions of political discourse and discussions taking place on social media, which also allows us to answer RQ2 -- whether a specific part of the platform's digital environment, its available reaction mechanisms, \textcolor{black}{relate} to the prevalence of the above political discourse types. 

\section{Answering RQ 2: Are different reaction mechanisms (i.e., upvotes, downvotes, no votes) associated with  different rhetoric components of political discourse in political discussions on Reddit?}

\subsection{Difference in Differences Analysis (DID)}

On Reddit, subreddit moderators are free to customize the user interface. In particular,  moderators can determine the types of reaction mechanisms (upvote, downvote) that subreddit members can use for interacting with other members. This creates a rich pool of behavioral data that describe political discourse dynamics in the presence or absence of different reaction mechanisms. To assess \textcolor{black}{this relationship}, we created \textcolor{black}{a quantitative model} based on a difference in differences (DID) analysis that compares how specific interventions, i.e., the change of available reaction mechanisms by moderators within subreddits, \textcolor{black}{relate to changes in} the type of political discourse  among users. \textcolor{black}{In general, DID analysis attempts to measure the effects of a sudden change in the environment, policy, or general treatment on a group of individuals or entities \cite{goodman2021difference}. It evaluates how the time-series or observations of a treatment group suddenly change based on a specific intervention, compared to a control group that is not subjected to the treatment. DID largely assumes that in the absence of treatment, the average outcomes for treated and comparison groups would have followed parallel paths over time, without any significant variation \cite{callaway2021difference}. Therefore, any difference between treatment and control after the intervention can be attributed to the intervention itself, revealing causal relations. In our case, although we apply DID and fulfill the parallel-trends assumption in pre-treatment periods, we are still careful in reporting our results, which we claim are mainly observational. Detected associations related only to the specific social media platform and time-periods, and to generate generalized knowledge, more in detail experimentation and research studies need to take place.}

\vspace{2mm}

To evaluate changes in reaction mechanisms, we selected posts from ``treatment'' subreddits that underwent a change in a reaction mechanism. We defined the ``baseline'' period as the period of time in which the subreddit operated with the default reaction mechanism (i.e., both up and down votes were available) and the ``intervention'' period as the period of time after the change in reaction mechanism was implemented. In our data, all subreddits started with the same default reaction mechanisms. Then,  moderators had the option to change the reaction mechanism. As a result, any change in political discourse between the baseline and intervention period can reflect a maturation effect  ~\cite{campbell2015experimental} rather than an effect of the intervention. To account for this possibility (and besides controlling for time), for each treatment subreddit, we also created a matched ``control'' sample from a subreddit \textit{that did not undergo a change in reaction mechanism} but that otherwise had the same characteristics as the treatment subreddits during their baseline periods. 

\vspace{2mm}

 We first selected all posts from subreddits that did not undergo a mechanism change but that were posted within the baseline or intervention period for each individual treatment subreddit. By aligning posts made during the same time window for the treatment and control subreddits, we could control for any overarching maturation effects that might have similarly affected both posts in the treatment and control subreddits (e.g., changes in world politics). We then calculated the Pearson correlation between the average discourse elements' scores for posts made during each treatment subreddits’ baseline period and the average discourse elements' scores for posts from each potential control subreddit made during that same period. For each treatment subreddit, we then selected the control subreddit for sampling that had: 1) a large number of posts during both the baseline and intervention period for the matched treatment subreddit; and 2) a high correlation between average discourse elements' scores during each treatment subreddit's baseline period. \textcolor{black}{ A high correlation coefficient between control and treatment subreddits ensured that the ``parallel trend'' assumption of the DID design  was fulfilled.} The extracted treatment  and control subreddits are presented in Table \ref{treatment}\textcolor{black}{, together with exemplary diagnostics plots in Appendix \ref{did_appendix}. These plots verified that the matched time-series were similar in levels and in trends, as advised by Kahn-Lang \& Lang  ~\cite{kahn2020promise}, who argued for a more careful election of control and treatment groups}. 

\vspace{2mm}

To further control for other factors that could plausibly affect the outcome, we again used the Wayback Machine  ~\cite{internetarchive} to extract different moderation rules that existed in each subreddit during the investigated period. Since we analyzed how people discuss within each community, we controlled for further factors that could have influenced interactions  ~\cite{perrault2019effects,andalibi2016understanding}. We create six different variables that represent different moderation rules that could be associated with how user discourse takes place  ~\cite{matias2019preventing}: \textit{Anonymity}, which describes whether a user's real identity should remain hidden in a subreddit. \textit{No troll}, which encompasses guidelines that explicitly prohibit trolling/spamming behavior that in themselves contain the usage of inflammatory language or repeated posting of nonconstructive information that can make discourse less deliberative. \textit{No hate-speech}, which forbids the usage of offensive and hateful speech towards individuals and social groups, which generally leads to emotional and ungrounded rhetoric. \textit{Civility}, which encompasses direct prompts in the guidelines to use civil language, and \textit{deliberation}, which includes moderation rules that promote evidence-based arguments and multi-perspective discussions. We also create a variable \textit{in-group} for subreddits that do focus only on the perspective of one social group (e.g., r/vegan, r/enoughtrumpspam) and explicitly mention in their guidelines that other opinions about an issue will not be tolerated, potentially leading to higher ``we vs. them'' rhetoric and less counterargument structures. \textcolor{black}{These variables serve as proxies of either the implementation of rules that explicitly aimed to alter the nature of discourse  ~\cite{saha2020understanding,birman2018moderation,dosono2019moderation} or that have been implemented because of abrupt incidents and patterns in user dynamics that moderators wanted to control  ~\cite{thach2022visible,squirrell2019platform,kiene2016surviving}}. Besides these moderation rules, we also use the ``nest level'' of a comment as a control. The nest level quantifies how deep a comment appeared in a specific discussion.  

\vspace{2mm}

We detected three main interventions in subreddits. The first one encompasses a set of subreddits that changed the baseline reaction mechanisms (up/down votes) to ``only upvoting'' (intervention A), a further set of subreddits that changed ``only upvoting'' to ``no voting'' (absence of available reactions, intervention B), and a set of subreddits that directly changed from the baseline to ``no voting'' (intervention C). Therefore, we created three different DID models that included the ``treatment'' subreddits together with the matched ``control'' subreddits that had the general form:

\begin{equation}
Prevalence_{c,n,d,s,i, year} = b_{n,d} + b_{d,s} + b_{i,d} +b_{d,year}+ \sum^6_1{b_{m,d}}
\end{equation}

, where \textit{$Prevalence_{c,n,d,s,i, year}$} is the value of the latent variable for discourse \textit{d} as predicted by the best CFA model for a specific comment \textit{c} that belongs to subreddit \textit{s}, has the nest level \textit{n}, given the presence (absence) of intervention \textit{i} at a specific year.  \textit{$b_{d,s}$} corresponds to the intercept for each discourse \textit{d} at subreddit \textit{s}, \textit{$b_{i,d}$} \textcolor{black}{the relationship size between intervention \textit{i} and discourse type} \textit{s}, \textit{$b_{n,d}$} the relationship between a comment's nest level and the prevalence of discourse \textit{d}, \textit{$b_{d,year}$} the general discourse prevalence at a specific year, and \textit{$b_{m,d}$} the estimator describing the association between the six moderation types \textit{m} and each discourse \textit{s}. For each of the three models, \textit{$i=1$} represents a change of the corresponding reaction mechanisms for the ``treatment'' subreddits.

\vspace{2mm}

\textcolor{black}{ 
This model structure is essentially based on Chiou et al.  ~\cite{chiou2018fake} for measuring effects between social media platform interventions on advertising and user sharing behavior. It allows us to take the subreddit's discourse heterogeneity into account}, ensuring that the results are not driven by the subreddit with the largest number of observations, while \textcolor{black}{controlling for each subreddit's specific effects and time effects}. \textcolor{black}{In our model, we exploit that the change in reaction mechanisms was not caused by how user discussions in terms of their political discourse took place. Reaction design interventions were related to how users misused the buttons (i.e., they used them to declare content preference over content fit in the subreddit, violating the unofficial reddit rules)  ~\cite{yarosh_2017}. They also corresponded to moderators' theoretic conceptions about whether the logic of up- or downvoting conformed to the scope of the subreddit  ~\cite{samineru}, hence being unrelated to whether users generated language containing specific rhetoric  components.} Overall, the model's focus on interventions A (up/downvotes to only upvotes), B (upvotes to no votes) and C (up/downvotes to no votes) included 48, 65, and 13 million observations respectively. Given the large number of observations, we created 100 mutually exclusive stratified samples for each case, on which we ran the models, and by bootstrapping calculated the corresponding mean values. \textcolor{black}{We ran models on a random sample of 1\% of the observations, and extracted clustered standard errors by subreddit, in order to avoid issues caused by the non-independence of observations within each subreddit (see e.g.,  ~\cite{bertrand2004much})}.

\begin{figure}[htp]

\subfloat{%
  \includegraphics[clip,width=1\columnwidth]{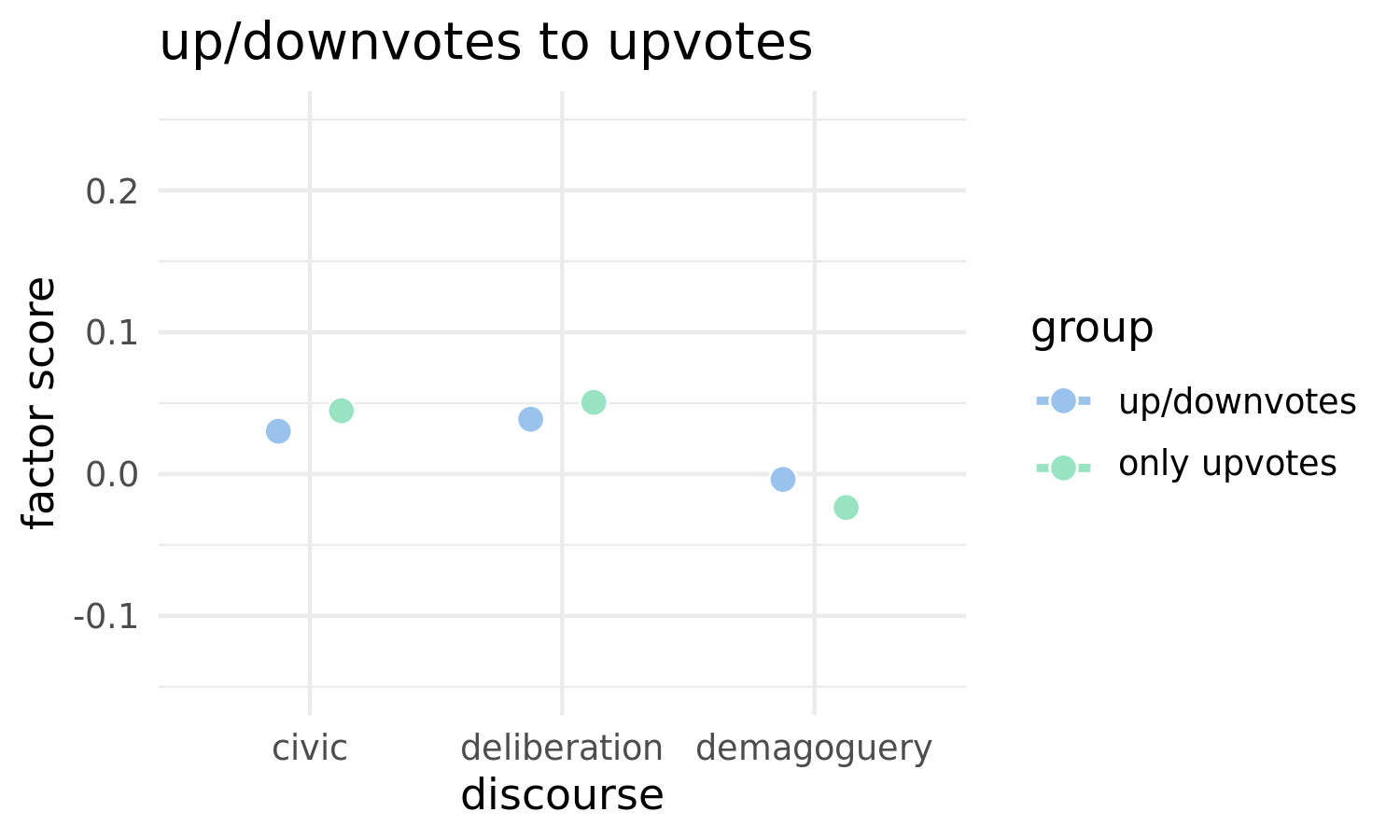}%
}
\vspace{2mm}
\subfloat{%
  \includegraphics[clip,width=1\columnwidth]{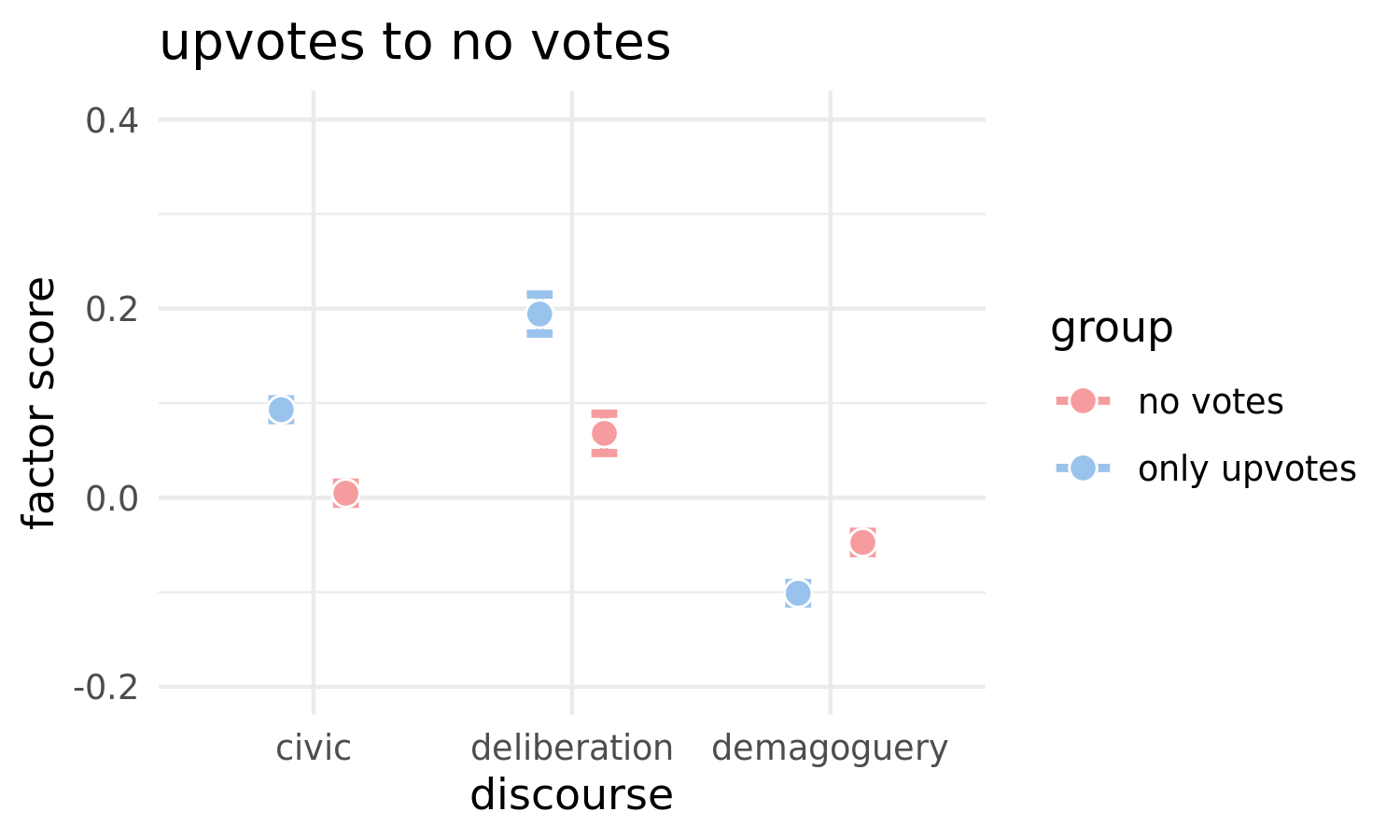}%
}
\vspace{2mm}
\subfloat{%
  \includegraphics[clip,width=1\columnwidth]{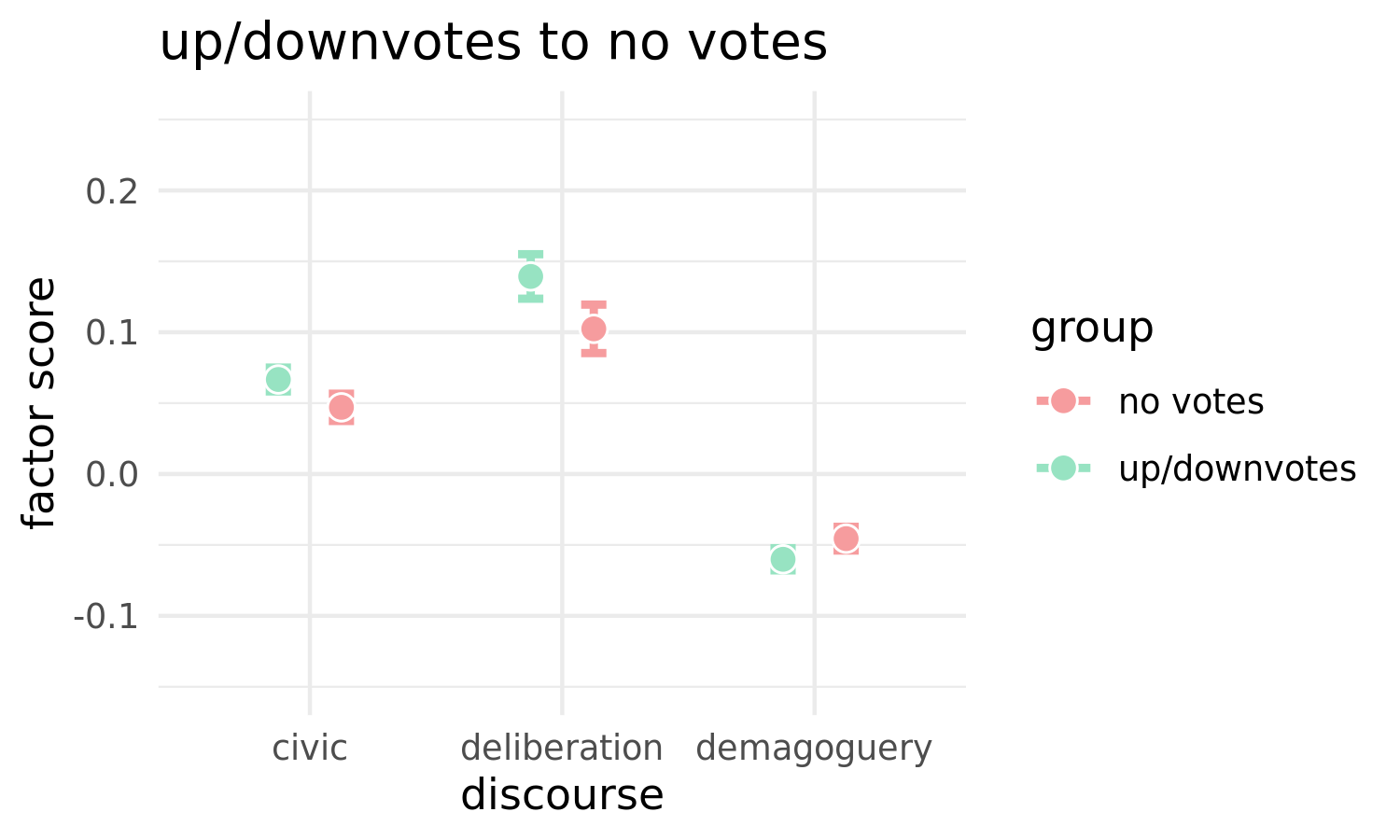}%
}
\caption{\textcolor{black}{Relationship of three interventions with} the prevalence of civic, deliberative, and demagogic discourse: (a) change from up/downvotes to only upvotes, (b) change from only upvotes to no votes, and  (c) change from up/downvotes to no votes. The y-axis describes the mean value of each factor in the specific subreddit sample. Values are not comparable across factors or interventions. Figure (a) also includes standard errors, but are too small to appear in the figure.}
\label{reactions_}

\end{figure}

\subsection{Results for RQ2}

The DID analysis showed that different reaction mechanisms corresponded to different forms of political discourse that users engaged with. Here, two findings are particularly relevant: First, our analysis demonstrates that subreddits with upvoting only had an increased prevalence of deliberative and civic discourse. Second, we find that no voting at all, the complete absence of reaction mechanisms, was associated with the highest levels of demagogic rhetoric. 

\vspace{2mm}

Figure \ref{reactions_} presents how changes in reaction mechanisms \textcolor{black}{related to} civic, demagogic, and deliberative discourse in each of the three distinct cases. Since we used the predicted scores for each discourse by the CFA model, which set factor variance to unity, the results are not comparable between discourses and interventions. Instead, we can assess the relative change within a discourse, given factor scores before and after an intervention took place. 

\vspace{2mm}

For subreddits where moderators decided to hide downvoting but keep upvoting (r/Conservative, r/EnoughTrumpSpam, \seqsplit{r/GenderCritical}, r/atheism, r/politics, r/exmuslim, r/ukpolitics), the prevalence of civic and deliberative discourse increased on average 0.47 and 0.30 times respectively, while demagogic discourse decreased 1.68 times. \textcolor{black}{For example, unsupported arguments on r/politics  decreased from 11\% to 9\%, while fact-related and structured arguments increased from 47\% to 49\%.} Removal of downvoting \textcolor{black}{was associated with} a significant reduction of demagogic rhetoric. An inverse \textcolor{black}{relationship} was observed when a subset of these subreddits subsequently decided to further remove the upvoting mechanism as well (r/Conservative, r/EnoughTrumpSpam, r/GenderCritical, r/atheism, r/politics, r/ exmuslim). In this case, civic and deliberative discourse decreased 0.94 times and 0.64 times respectively, while demagogic discourse increased 0.53 times. \textcolor{black}{For example, the discourse on r/Conservative contained 23\% more arguments that belong to demagoguery (13\% with only upvote, 16\% with no reaction mechanism). In contrast, \textit{collective rhetoric}, which belongs to civic discourse, decreased from 3.3\% to 3.1\%.} 

 In subreddit interfaces without reaction mechanisms, the discourse became significantly more demagogic compared to an environment with only upvoting. Moreover, subreddits who immediately removed all available voting mechanisms from the baseline state (r/unpopularopinion, r/vegan), encountered a decrease in civic and deliberative discourse by 0.27 and 0.34 times, respectively. They also faced an increase in demagogic discourse by 0.22 times. Again, comparing up and downvoting to no voting in subreddits, no voting \textcolor{black}{was associated with} a significantly lower level of constructive dialogue. \textcolor{black}{For example, the discourse on r/politics contained 15\% less fact-related and structured arguments (47\% of comments contained one of the two arguments when both votes were available, while only 40\% after). In contrast, the usage of unsupported arguments increased by 17\% (5.8\% with both reaction mechanisms, 7\% with no reaction mechanism). These absolute values in the percentage changes of specific arguments provide a better understanding about variations in the discourse taking place, albeit are not controlled for time and community-specific effects, which the DID model actually accounts for. For example, there was a declining trend in deliberative discourse over time across all subreddits, which partially masks the difference between the levels of deliberative discourse in environments having up/downvotes and only upvotes, since the change from up/downvotes to another reaction mechanism structure always appeared later in time.}

\vspace{2mm}

These results offer a clear answer to RQ2: Changes in reaction mechanisms \textcolor{black}{corresponded} to changes in the nature of the political discourse taking place in political subreddits between 2010 and 2018. Only upvoting \textcolor{black}{was associated with} more civic and deliberative discourse, while the presence of downvoting \textcolor{black}{was associated with a decrease in} their prevalence. Furthermore, the complete absence of any voting means \textcolor{black}{was associated with} the lowest level of civic and deliberative discourse. We observed an inverse behavior for demagoguery, with the upvoting mechanism being associated least with the discourse, while downvotes and no-votes \textcolor{black}{were associated with} significantly increased demagogic rhetoric. Although our analysis did not focus on understanding why these effects appeared on Reddit discussions, our findings correspond to previous studies that show that the absence of negative reinforcement (downvoting) improved the quality of discussions  ~\cite{khern2020haters}, but also that the type and magnitude of feedback users receive  does influence the future content they will generate  ~\cite{adelani2020estimating}.

\section{Discussion}

Our analyses shed light on how users discuss political topics on Reddit through the lens of three prominent political discourse theories. When using deliberative rhetoric, interactions were fact-related and structured and lacked statements without evidence. In contrast, they did not recognize the perspectives of other authors (empathy/reciprocity) or provide counterarguments. 

When the discourse contained civic features,  users underlined a collective identity and made calls for situational actions. Nevertheless, there was no significant use of emotional or unstructured/nonfactual language as suggested by civic discourse scholars. 

When the discourse was demagogic, users mostly engaged in emotional language and nonfactual arguments, but they did not make calls for generalized arguments or focus on the importance of the ``people'' as suggested by theorists. These elements of discourse show both that discussions on Reddit  differ from theoretical conceptions and from how these discourse types might be deployed in other environments or by other discussants. For example, politicians on social media, when creating demagogic statements, often use ``you in the epicenter'' arguments  ~\cite{bobba2019social}, which we did not find for Reddit users in our analysis. 

Focusing on the overlap between theories and empirical data, we detected core elements of the theories in our sample discussions and used them to evaluate political discourse. Simultaneously, we located specific divergences that might create concerns as to whether normative theoretic conceptions can actually be translated into discursive practices. These questions necessitate a further systematic analysis and measurement of theories in different environments and conditions, which were not performed by us. Nonetheless, our study emphasizes the relation between theory and political discussions on social media. It further serves as a proof-of-concept on how to measure political communication from the lens of political theories and recursively verify, falsify, or reevaluate the use of existing political theories when evaluating political communication. We believe that this is a valuable contribution in the field, since the relationship between social media and democracy and its core values is under heightened scrutiny. For example, our study informs discussions in political discourse theory that seek to understand whether political discourse on social media is essentially distinct and unique in contrast to political discussions "offline". As such, our findings could inform scholarly discussion on political discourse theory. 

Moreover, our study produces an understanding of the \textcolor{black}{relationship between different reaction mechanisms and} the type of political discourse user engage in. We  found that having ``only upvotes'' as reaction mechanism was the most beneficial for the prevalence of deliberative and civic discourse, while the absence of voting favored the generation of discussions with more demagogic rhetoric. These results contribute to an ongoing debate on how to design social media platforms. Recently, YouTube experimented with the removal of downvoting from its platform  ~\cite{youtube_2021}, while Twitter recently incorporated downvoting in its user interface  ~\cite{hunter_2022}.

Based on our findings, it seems that up- and downvoting \textcolor{black}{is associated with} an increase of demagogic discourse compared to only upvoting, and we hope that our findings can be taken into consideration when making such decisions in the future. Although we controlled for multiple factors and based our modeling decisions on prior scientific work, additional experiments and studies are needed to further validate whether results support the existence of a causal relationship. A reaction mechanism can \textcolor{black}{relate to} discussions in different ways, many of which we did not analyze in our study (e.g., prevalence of hate speech \cite{sasse2022prosocial, cyprisintervening} or user participation \cite{engelmann2018democracy}). Reaction mechanisms can also \textcolor{black}{play a variable role} given different user demographics and the nature of the platform. Hence, we do not claim that our results are necessarily generalizable, and we argue that further research studies should continue to systematize knowledge to provide policy recommendations for designing social media platforms that promote democratic values. 

\textcolor{black}{As we pointed out in Section \ref{affordances}, our investigation concentrated on \textit{reaction mechanisms} rather than \textit{affordances}. Future research should further investigate the relationship between community-specific uses of reaction mechanisms (i.e., as affordances) and political rhetoric among social media users.} \textcolor{black}{Users do not evaluate reaction mechanisms simply by their technical functionality. In Reddit, voting is inseparable from the culture of the platform, with users disregarding platform rules by making, and enforcing, their own rules and norms as voting is used for pointing out who and what is ``right'', to assign and recognize social status, and to negotiate meaning and ethical norms \cite{graham2021sociomateriality}. While reaction mechanisms create the propensity to behave in specific ways, what kind of behaviors will dominate interactions depends on complex social dynamics in online communities that are difficult to control for. It is no coincidence therefore that the same reaction mechanisms can have varying effects in different digital environments and also can lead to different dimensions of user behavior (e.g., the content of posts, the frequency of posting, how long users will remain in a community, etc.)  \cite{cheng2014community,khern2020haters}. This social dimension of reaction mechanisms can already be recognized when evaluating the reasons why moderators on specific subreddits decide to change reaction mechanisms. For example, in our own analysis, we found that the \textit{exmuslim} subreddit deactivated downvotes because of button abuse and because its usage did not end up conforming to the general rules of reddit, while other subreddits did not have a downvote in order to create a ``safe space'' for its members. These discrepancies reveal the relationship between social dimensions of the platforms and their technical design, which cannot be neglected when integrating or evaluating design features on platforms.}

\textcolor{black}{Building on this sociotechnical perspective}, future research should focus on understanding \textit{why} reaction mechanisms \textcolor{black}{relate in specific ways to positive and negative feedback from a democratic perspective} and how this can be integrated ideally in platform design. \textcolor{black}{Properties such as platform objective (political vs. non-political topics, humor vs. deliberation) and user group properties (e.g., age, gender, political background), can interact in unforeseen ways with design features, affecting user discussions and behaviors (see for example political identity induced differences in content moderation effects \cite{papakyriakopoulos2022impact}). Furthermore, what is democratically valuable is contestable, and even the evaluation about specific effects of reaction mechanisms is not clear cut. For example, expressing emotion might be detrimental for deliberative discourse, but empowering from a civic perspective; therefore, an increase or decrease because of technical design might be favored in some cases and not in others}. Especially since reaction mechanisms are fundamental data inputs for recommendation systems that distribute and shape the flow of information between users, it is important to quantify biases, issues, and effects that related  reaction mechanisms bring to the digital ecosystems. \textcolor{black}{We argue that the proposed methodologies in this paper can contribute towards that end as they are able to directly associate political theories to empirical user behavior, which can inform policy decisions based on more scientifically and theoretically grounded democratic frameworks. Furthermore, we hope that researchers advance our research project and its proposed methodologies to better understand combining machine learning tools with classical statistics.}


\section{Conclusion}

Our study introduced a framework for evaluating the empirical manifestation of political theories on social media. We performed a comparative analysis of three prominent theories of political discourse, i.e., deliberative, civic, and demagogic. We produced a set of constitutive rhetoric elements that we referred to as a ``minimal conceptualization'' of these three discourse theories. We then used multilabel classification to explore the extent to which these rhetoric  components  characterize 155 million user comments across 55 political message boards on Reddit. We find that essential components of the three discourse theories indeed characterize political user discussion. Nonetheless, we also found that specific theoretically defined elements of discourses did not resurface in the political discussions on Reddit. Over a time span of eight years (2010-2018), we created a \textcolor{black}{quantitative setup} to identify changes in political discourse as a function of introducing or removing upvotes, downvotes or both. We showed that social media reaction changes were associated with changes in the nature of political discourse. Interfaces with upvoting only were associated with the highest level of civic and deliberative discourse, while the absence of any reaction mechanisms showed the strongest manifestation of demagogic rhetoric. We believe that these results are valuable contributions to ongoing policy and research discussions on platform design and its important role in supporting more civil interaction on social media. 

\begin{acks}
This study was funded in part by the Princeton Center for Information Technology Policy and supported by a Princeton Data Driven Social Science Initiative Grant. We gratefully acknowledge further financial support from the Schmidt DataX Fund at Princeton University made possible through a major gift from the Schmidt Futures Foundation. We  thank Arvind Narayanan for his conceptual support in the first stages of the study, Aaron Snoswell and Oscar Torres-Reyna for their methodological advice, and Andrés Monroy-Hernández, Jens Grossklags \& Michael Zoorob for their feedback on the final manuscript. We are further grateful for the constructive feedback received from the Princeton Center for Information Technology fellows meeting and the Work in Progress reading group, and the MPSA 2022 Panel on  Social Media Ecosystems and Their Effects Across Platforms. 
\end{acks}
\bibliographystyle{ACM-Reference-Format}
\bibliography{main}

\appendix
\onecolumn


\section{Ethical Concerns}

The collected Reddit data are public data. Nevertheless, authors have raised concerns about the privacy, anonymity, and discoverability of individuals included in them  ~\cite{proferes2021studying}. Hence, we performed further actions to ensure the ethically just processing of user comments, taking also into consideration the guidelines of the association of internet researchers  ~\cite{franzke3association}. First, we stored our data on a safe, password protected server. We plan to delete individual observations upon publication of the study and only keep necessary aggregate materials to replicate findings. Furthermore, we included in the appendix some example user comments, in order to explain our methodology. We ensured that someone cannot match these comments to the accounts that generated them on Reddit, either by using search engines such as Google, or the Reddit search function, protecting user privacy to the maximum possible degree. 

\section{Further explanation of labels with example comments}  \label{definitions_appendix} 

\subsection{Unsupported argument}

We developed a label to account for social media comments that made a proposition without providing any supporting justifications: a label called “unsupported argument” was used to mark comments that contained a proposition with a pragmatic purpose (for example, to underscore political identity/group identity) but lacked evidence to support the empirical validity of the overall proposition. We also applied this label to arguments that included some supporting reasoning, which was, however, not relevant for the proposition. Thus, unsupported arguments were instrumental in arguing for a position (within a given context) while providing evidence that was not relevant for the argument’s context. 

\vspace{2mm}
Examples of unsupported arguments: 

\begin{itemize}
 \item \textit{``Democrats shun businessmen for making profits that are reinvesting back into the economy instead of letting the govt. tax those profits more, meanwhile the first lady models \$40k bracelets?? Class warfare...''}
\vspace{2mm}
 \item \textit{``Not at all surprised. Isn't the conservative answer to everything 'more guns, more violence'?''}
\end{itemize}
\vspace{2mm}
We used the label ``unsupported argument'' as characteristic of demagoguery, irrelevant for civic engagement, and not characteristic of deliberative discourse. 

\subsection{Structured argument}

Second, we used one label to analyze comments at the level of the syntax only. We called this label ``structured argument.'' A comment was labeled as a structured argument when it consisted of syntactically logical structure: a proposition with a justification connected by coherent sentence structures. 

\vspace{2mm}
Here is an example of a comment that we annotated the label structured argument:
\vspace{2mm}
\begin{itemize}
 \item \textit{``Libertarians aren't big fans of militant interventionism. Frankly, I don't know why anyone would be, considering our dismal record of 'nation building' and the fact that we're going broke trying to police the world.''} 
\end{itemize}
\vspace{2mm}
We used the label “structured argument” as not characteristic of demagoguery and civic engagement, and characteristic of deliberative discourse. 

\subsection{Fact-related argument}
\label{fact_related}

Third, we developed a label called ``fact-related argument.'' Given that we could not falsify a comment's truthfulness, we called the label fact-\textit{related} argument, which classified comments according to one of two types of justification: empirical justifications and reasoned justifications. When supporting a claim, empirical justifications provided either a direct reference to other sources (e.g., in the form of links or article references) or referred to personal experiences and anecdotes relevant for the overall claim. For example, comments that presented statistical information about unemployment rates together with a reference to external sources were labeled as fact-related argument.

\vspace{2mm}

\begin{itemize}
\item \textit{``It means when a business will enter the market it won't act differently than other businesses. In this case meaning paying their workers higher wages. The environment of running a business means you want to keep costs down to maximize profit, this means that businesses will follow this principle. Businesses in similar workforce environments will all settle on extremely similar wage settings because of isomorphism. (your workforce population being the outside environmental factor, which is the same to all the businesses in the same region as the workforce). \url{http://en.wikipedia.org/wiki/Isomorphism_\%28sociology\%29”}''}
\vspace{2mm}
\item \textit{``U6 - The true measure of the state of Unemployment remained unchanged at 14.7\%.''}
\vspace{2mm}

Moreover, the comment below also described unemployment benefits. It contained a proposition backed up by personal experience and anecdote:  
\vspace{2mm}
\item \textit{``This is the problem though isn't it. I mean, I worked from 14 (whilst at school - Woolworths and then a factory) onwards. I didn't go to college or university because I couldn't afford to (and instead paid for others to go..) but did perfectly well without it. I haven't claimed unemployment benefits…''}
\vspace{2mm}

In contrast to empirical justification, reasoned justification did not refer to empirical references. Reasoned justification referred to comments that included a proposition justified by reasons that fell back on conventional and common sense knowledge that were relevant for the overall proposition (i.e., had a pragmatic outlook). 
The following Reddit comment from the Subreddit ``Birthcontrol'' is an example of comment we labeled as fact-related because of its reasoned justification:
\vspace{2mm}
\item \textit{``I don't know specifically how much each type costs (there are apparently fucktons of different female birth controls, and I'm a dude), but isn't it naiive to think that just because someone can't afford BC that they'll abstain from sex? That's similar to the line of thinking people use when they advocate abstinance-only sex ed. The places that use this tend to have the highest levels of teen pregnancy, because in the end, these are very natural urges for people to have, and to expect that people will refrain from doing it just isn't realistic or practical. Providing options to make it safe and preventative simply works best.''}

\vspace{2mm}

We used the label ``fact-related argument'' as characteristic for deliberative discourse and as not characteristic of civic engagement and demagoguery. 
\end{itemize}

\textcolor{black}{Challenges of defining fact-related argumentation among co-authors: Authors agreed that the rhetoric component ``fact-related argument'' should be assigned to deliberative discourse. Initially, however, authors did not agree on the definitional scope of fact-related argument. One author argued for a ``strict'' definition that would label user comments as fact-related only if they provided a verifiable empirical justifications for a claim in the form of a link to an external source (see, empirical justification above). After further engagement with the literature on deliberative discourse, it became clear, however, that such a strict requirement for the truthfulness of a claim was not supported by the literature on deliberative discourse (see, for example,  ~\cite{sep-habermas, dahlgren2006civic, dahlgrenonline}). Instead, scholars on deliberative discourse state that it emerges in a communal setting where participants exchange ideas based on relevant experiences that they have made in relation to a communal issue or challenge. Thus, we extended the definitional scope of fact-related argument to comments that referred: a) to personal experiences or anecdotes, and b) that included arguments justified by reasons that were relevant for the overall proposition (see, reasoned justification above). Arguments that were neither backed up by external sources or personal experiences and that provided arguments that were not relevant for a proposition were labeled as ``unsupported argument.'' Moreover, definitional work on fact-related argument led one author to propose a label that would classify comments according to their syntactical logic, which resulted in the development of the label ``structured argument''. Thus, discussions among co-authors on the definition of fact-related argument was productive and led to further definitional differentiations.}

\vspace{2mm}

\textcolor{black}{There was initial disagreement on the assignment of fact-related argument to civic discourse. Some scholars (e.g.,  ~\cite{mccoy2002deliberative}) argue that deliberative discourse is instrumental for the development of a common policy goal. Structured interaction settings support the development of a common goal among participants. However, we decided against including fact-related argument in civic discourse. First, social media platforms are not designed for the purpose of defining a common policy goal among community members. Second, a large share of scholars (see Section \ref{civic}) underlines the importance of inclusive interactions that put little conversational constraint on participants.}

\subsection{Counterargument}

Fourth, we called one label ``counterargument''. We applied this label to comments that included contradicting responses to previous comments. For a response to qualify as counter argumentation, responses had to either make a clarifying statement to resolve misunderstandings or misinterpretations of a previous comment or contain arguments against a proposition made in another comment. Counter arguments could aim at the overall proposition of a comment, a single proposition (if multiple propositions were present), the reasoning behind a proposition, or the evidence that was brought forward to support a proposition. Generally, only comments that responded to other users' comments in a constructive manner were labeled as counterargument. We labeled comments as counterargument only when they allowed for or even invited further discussion on a topic. Comments that included strong insults were not considered for this classification. 
\vspace{2mm}

An example comment for a \textit{clarifying} counterargument is: 
\vspace{2mm}

\begin{itemize}
\item \textit{``Never said the article's proposed solution was conservative. 
I didn't say that. I said sentiments are not the same as arguments. The support of sentiments for the sake of supporting sentiments is one of the stupider things I've heard in my lifetime. Sorry to say. Not that I think that reflects on you in anyway.''}
\end{itemize}
\vspace{2mm}

An example comment for a \textit{contradicting} counterargument is: 
\vspace{2mm}
\begin{itemize}
\item \textit{``Nope. I think they were definitely abusing their power and went too far.
Now the question is... Was the response to the trial fair? Hmmm did normal citizens who had nothing to do with the police beating of Rodney King deserve their homes and businesses vandalized, looted, and burned? Nope''}
\vspace{2mm}
\end{itemize}
We used the label ``counterargument'' as characteristic of civic engagement and deliberative discourse and as not characteristic of demagoguery. 

\subsection{Empathy/Reciprocity}

We further created a label to classify comments that recognized and acknowledged another user's proposition. Here, we classified comments that indicated a user’s perspective-taking toward a previously made comment by another user. Such comments could extend or add to the propositions made by other comments and did so in a constructive and non-hostile manner. 
\vspace{2mm}

Two examples of comment labeled as ``empathy/reciprocity'':
\vspace{2mm}
\begin{itemize}

\item \textit{``You do realise that the NHS already makes those kind of decisions? People are woefully unaware of the limitations of the NHS. Indeed, but on the basis of need and cost as a whole, not personal wealth, which is an important distinction (probably *the* important distinction), not to mention that an insurer would too, but quite possibly with a different set of incentives. It's one of the reasons I don't like the marketisation of the NHS, its not the right focus for a public health provider.'' }
    
\vspace{2mm}
    
\item \textit{``I agree, but then where that line is drawn maybe the battle ground. Exactly, but we need to be honest in the discussion, rather than using edge cases that distort the reality of the systems we are aiming to improve.''}
\end{itemize}
\vspace{2mm}
We used the label empathy/reciprocity as characteristic of civic engagement and deliberative discourse, and as not characteristic of demagoguery. 

\subsection{Emotional language}

We used one label to classify comments that contained emotional language. The text contains (positive or negative) sentiment related adjectives, swear words, offensive, aggressive, or satirical language. It also includes syntactical features in the text that are indicative of emotionally-laden speech such as the use of Caps Lock.
\vspace{2mm}
Examples of two comments that used emotional language:
\vspace{2mm}
\begin{itemize}
\item \textit{``What the fuck is this ceremony? Bowing every 5 steps? Loyal to the Queen? Jesus Christ this is an antique piece of shit we need to do away with.''}
\vspace{2mm}
\item \textit{``They both are a threat to our nation and bastardize our democracy... Sounds like you buy into one side's rhetoric.''}
\end{itemize}
\vspace{2mm}
We used the label emotional language as characteristic of demagogic and civic engagement, and as not characteristic for deliberative discourse.

\subsection{Labels that signify a ``call for action''}

We developed two labels that accounted for comments that included a ``call for action.'' These labels allowed us to identify whether individual comments promote behaviors that are associated with demagoguery, civic engagement and deliberative discourse. We named one label ``generalized call for action'' and another label ``situational call for action.'' 

\subsubsection{Generalized call for action}

Generalized calls for action communicated a need for change in the context of the discussion topic. Comments called people to act or deal with a topic in a general manner, without providing any explicit details on what kinds of actions this would require and how this could be done. The need for a particular action was not justified by reference to any form of evidence. Generalized calls for actions are essentially ``empty'' messages promoting social change (broadly defined). They include arguments about how society should be, without stating why it should be so, or how someone can realistically achieve such goals.
\vspace{2mm}

Example comment labeled as generalized call for action:
\vspace{2mm}
\begin{itemize}
 \item \textit{``.... So we need to find a way to rebuild the link in peoples heads between what they put in and what they get out, so we don't have people sitting on houses worth £500K expecting the state to pay for their end of life care.''}
\vspace{2mm}
\end{itemize}
We used the label generalized call for action as characteristic of demagoguery and as not characteristic for civic engagement and deliberative discourse. 
 
\subsubsection{Situational call for action}

In contrast to the concept ``generalized call for action'', ``situational call for action'' includes comments that promote a specific public goal and/or describe, at least to some degree, how a public goal can be achieved. Here, comments call for action based on facts and evidence and refer to a specific situation, a case study, a location, time, or legislation. Arguments related to situational calls provide concrete information about how to organize and collaborate in the real world, what actions should be taken and towards what purpose. 
\vspace{2mm}

Example comment labeled as situational call for action:
\vspace{2mm}
\begin{itemize}
 \item \textit{``...Use checks only, get off of social media or anything that tells people where you are and what you are doing, use cash, avoid online payments (send a check), etc. It's just easier to let your self be a part of all of that but there are ways to get around it, but like I said it's just easier to be a part of it all. Sure you can't avoid paying taxes, having a SSN and using it, and giving people your information for work purposes, but you can certainly so things to leave less of a trail of your self that's all the easier to be looked at. I know I sound very paranoid, but as I stated before, Orson Wells was half right, the difference is we put our seleves out there.''}
\vspace{2mm}
\end{itemize}
We used the label ``situational call for action'' as characteristic of civic discourse and as irrelevant of deliberative discourse and demogogic discourse.

\subsection{Identity labels}
\label{identity}

We developed four different labels that classified comments with references to identity. 

\subsubsection{Collective rhetoric}

Collective rhetoric focuses on the use of words such as ``we'', ``our'', or any statement that reveals and promotes group membership and empowerment. Here, users presupposed a shared knowledge of the collective that they understood themselves to be part of.
\vspace{2mm}

Example comments labeled as collective rhetoric:
\vspace{2mm}
\begin{itemize}
 \item \textit{``My point was we the people are paying tax revenue to deal with this, instead of collecting tax revenue.''}
 \vspace{2mm}
 \item \textit{``Our society has been shaped by our two-party system (the result of the way our Constitution was drawn up). Our politics are now devolved into a two-sided mudslinging party because of it....When we have fewer outlets in which to express our political opinion, we are pigeon-holed into one. And these two-pigeonholes are of a large ideological variety in terms of members, yet in reality their platform only covers a small ideological spectrum. This results in passionate dislike of the other party.''}
\vspace{2mm}
\end{itemize}
We used the label collective rhetoric as characteristic of civic engagement and demagoguery, and as irrelevant of deliberative discourse. 

\subsubsection{You in the epicenter}

Moreover, one label called ``you in the epicenter'' marks comments that focus on the identity of a specific social group and its privileged standing. Such comments put a social group under the limelight, stating why they were so important without adequate argumentation. They also include populist statements referring to ``the people'' in general.
\vspace{2mm}

Example comments labeled as you in the epicenter:
\vspace{2mm}
\begin{itemize}
 \item \textit{``Your opinion does not encompass, we the people. It only includes the low information liberals that think government control is a good thing.''}
 \vspace{2mm}
 \item \textit{``One of the main reasons I became a conservative was I believe that America was founded on the idea of small government kept in check by the people. That we the people need less regulations, less of our tax money going to politicians pockets, and absolutely NO government interference when it comes to gun rights. That's why I am conservative.''}
\vspace{2mm}
\end{itemize}
We used the label you in the epicenter as characteristic of demagoguery and as irrelevant for civic engagement and demagoguery. 

\subsubsection{We vs. them}

As an extension of the label ``you in the epicenter,'' we  developed the label ``we vs. them''. Here, comments referred to the identity of a group or collective and affirmed this identity through the explicit contrast or degradation of another group. In ``we vs. them,'' two different groups were put in competition or conflict with each other highlighting the superiority of one group over the other. 

\vspace{2mm}

Example comments for we vs. them:
\vspace{2mm}
\begin{itemize}
 \item \textit{``No they couldn't. MRAs didn't fight against feminism, in fact most MRAs were feminists. We only became anti-feminist once feminists started fighting against us instead of supporting us in our fight for equality.''}
 \vspace{2mm}
 \item \textit{``AOC wants us to have a corrupt system like South America where the top 0.1\% still control everything though the government and the 9.9\% get fucked out of everything they own. I'm in the top 10\% of income earners in the USA. It isn't very hard to get into. We are the ones who get fucked by these socialist policies.''}
\vspace{2mm}
\end{itemize}
We used the label we vs. them as characteristic of demagoguery and civic engagement and as not characteristic of deliberative discourse.

\subsubsection{Who instead of what}

The text bases its argument on who a person or social group is and not on their behavior or actions. It includes arguments that refer to individuals in a discrediting tone, without justifying why. They also justify specific events on ungrounded features on someone's character, without providing actual contextual information of what happened or why an individual did something.
\vspace{2mm}

Example comments for who instead of what:
\vspace{2mm}
\begin{itemize}
 \item \textit{``Call them what they are White Christian nationalist.''}
 \vspace{2mm}
 \item \textit{``Cruel and unusual. Republicans are so evil.''}
\vspace{2mm}
\end{itemize}
We used the label who instead of what as characteristic of demagoguery, as not characteristic of deliberative discourse and as irrelevant for civic engagement.

\vspace{2mm}

\textcolor{black}{Challenges of defining identity labels among co-authors: Social media posts that discuss politics commonly include references to groups and the membership in groups or collectives. Engaging with the literature on demagogic and civic discourse revealed identity labels to be relevant for these two types of discourse but not deliberative discourse (for demagogic discourse, see  ~\cite{roberts2020demagoguery, gustainis1990demagoguery, hogan2006demagoguery}; for civic discourse, see  ~\cite{adler2005we, hauser2004rhetorical, bonilla2015ferguson}). However, conceptualizing different identity references for demagogic and civic discourse proved to be challenging and authors disagreed on how to define identity labels for demagoguery and civic engagement. Out of all three political discourse theories, civic engagement is the most contested one. It allows for the most diverse definitions of what it's essential components are (e.g., compare  ~\cite{diller2001citizens} and  ~\cite{bonilla2015ferguson}). We first agreed on a basic identity label to identify comments that expressed the existence of collective or group (i.e., collective rhetoric). One author then proposed an identity label to highlight group \textit{differences}, a feature of both demagogic and civic discourse  ~\cite{dahlgren2006civic, hauser2004rhetorical, hogan2006demagoguery}. This label was called ``we vs. them'' to underline the competitive nature of interaction in both demagoguery and civic engagement. In demagogic discourse, members often degrade another group to establish a sense of unity. Here, a unity is formed around group characteristics and people that appear to best embody such characteristics. In short, narratives of unity are not formed around a serious discussion around policy-making to bring about political change but around important personas. This, one author pointed out could be a decisive conceptual distinction between identity in demagoguery and civic discourse. To account for the lack of engagement with policy-making topics to establish a sense of unity (as is the case for civic engagement), we finally agreed on two more identity labels for demagogic discourse: ``you in the epicenter'' and ``who instead of what.'' The first label highlights the importance of a group and why it should be treated preferentially while the second appeals to the characteristics of in-group or out-group members. In civic discourse, we added the labels ``empathy/reciprocity,'' ``counterargument'', and ``situational call to action'' that, together with ``we vs. them,'' would sample comments that expressed a sense of unity but that at the same time featured constructive and productive interactions on specific policy-making goals.}

\subsection{General information}

Comments that mentioned events or facts without making a specific argument or proposition were classified as ``general information.'' 
\vspace{2mm}
\begin{itemize}
 \item \textit{``For the curious: \$14.22 in 1890 = \$388.08 in 2015
\$2,069.90 in 1890 = \$56,489.79 in 2015''}
\vspace{2mm}
\end{itemize}

\subsection{Other}

All other comments were classified as ``other.'' This included deleted comments, nonsensical statements, sentences without any syntactical structure, questions that could not be contextualized, random statements that could not be contextualized, or single words.

\section{Simulating the covariance between a discourse theory and a non-frequent elementary argument.} \label{simulation}

Modes of discourse are complex phenomena, and through our minimal conceptualization we associated multiple items with each of them. Since CFA has not been used extensively for the purpose of measuring and quantifying political theories, we created a simulation to understand how the not-so-frequent -but existing- relationship of an item with a construct is represented in terms of correlation \& covariance. We generated a sample of 20 observations, in ten of which appears a specific factor. Then we created an item that appeared four times in an observation that the factor appeared in, and one time in an observation that it did not appear. This resulted in a correlation of $cor = 0.34$, and a covariance of only $cov = 0.078$. Since it is very rare that a user comment will include many types of argumentation in it, elements of a theory will appear sparsely in the observations. Thus, low covariance relations reported by the CFA are still theoretically valid, as long as the model satisfies the corresponding goodness of fit criteria. 

\newpage

\section{Tables}
\label{tables_appendix}
\begin{table}[h!]
\caption{Description \& examples of each label extracted from the minimal conceptualization of deliberative, civic, \& demagogic discourse. }
\label{definitions}
\resizebox{0.943\columnwidth}{!}{\begin{tabular}{c@{}c@{}c@{}}
Label                       & Description                                                                                                                                                                                                                                  & Example comment                                                                                                                                                                                                                                                                                                                                                                                                                                                                                                                                                                                                                                                                                                                                                                         \\ \hline
you in the epicenter        & \begin{tabular}[c]{@{}c@{}}The text puts a social group under the limelight, \\ saying why they are so important without adequate \\ argumentation. It also includes populist statement\\ referring to "the people" in general.\end{tabular} & \begin{tabular}[c]{@{}c@{}}“Your opinion does not encompass, we the people. \\ It only includes the low information liberals that think \\ government control is a good thing.”\end{tabular}                                                                                                                                                                                                                                                                                                                                                                                                                                                                                                                                                                                   \\  \hline
generalized call for action & \begin{tabular}[c]{@{}c@{}}The text calls people to act/ deal with \\ a topic in a general manner, without providing\\  explicit details of how this could be done.\end{tabular}                                                             & \begin{tabular}[c]{@{}c@{}}“.... So we need to find a way to rebuild the link \\ in peoples heads between what they put in\\  and what they get out, so we don't have people\\  sitting on houses worth £500K expecting the\\  state to pay for their end of life care.”\end{tabular}                                                                                                                                                                                                                                                                                                                                                                                                                                                                                         \\  \hline
situational call for action & \begin{tabular}[c]{@{}c@{}}The text calls for action based on facts \\ and evidence and refers to a specific situation/case\\  study/location/time/legislation.\end{tabular}                                                                 & \begin{tabular}[c]{@{}c@{}}“Use checks only, get off of social media or anything\\  that tells people where you are and\\  what you are doing, use cash, avoid\\  online payments (send a check), etc. It's\\  just easier to let your self be a part of all of that \\ but there are ways to get around it, but like I said it's just \\ easier to be a part of it all. Sure you can't avoid \\ paying taxes, having a SSN and using it, and \\ giving people your information for work purposes,\\  but you can certainly so things to leave less of\\  a trail of your self that's all the easier to be\\  looked at. I know I sound very paranoid, but as I stated\\  before, Orson Wells was half right, the difference\\  is we put our seleves out there.”\end{tabular} \\  \hline
who instead of what         & \begin{tabular}[c]{@{}c@{}}The text bases their argument on who a person\\  or social group is and not on their behaviour or actions.\end{tabular}                                                                                           & “Call them what they are White Cristian nationalist.”                                                                                                                                                                                                                                                                                                                                                                                                                                                                                                                                                                                                                                                                                                                         \\  \hline
fact-related comment        & The text refers to an actual event having taken place.                                                                                                                                                                                       & \begin{tabular}[c]{@{}c@{}}"The true measure of the state of Unemployment \\ remained unchanged at 14.7\%.”\end{tabular}                                                                                                                                                                                                                                                                                                                                                                                                                                                                                                                                                                                                                                                       \\  \hline
structured argument         & \begin{tabular}[c]{@{}c@{}}The text provides a set of propositions that\\  clearly explain how an argument is supported.\end{tabular}                                                                                                        & \begin{tabular}[c]{@{}c@{}}“Libertarians aren't big fans of militant interventionism.\\  Frankly, I don't know why anyone would be,\\  considering our dismal record of "nation building"\\  and the fact that we're going broke trying to police the world.”\end{tabular}                                                                                                                                                                                                                                                                                                                                                                                                                                                                                                     \\  \hline
counterargument             & \begin{tabular}[c]{@{}c@{}}The text provides a reply to an argument \\ in a constructive manner. \end{tabular}                                                                                                                                                                            & \begin{tabular}[c]{@{}c@{}}“Never said the article's proposed solution was conservative.\\ I didn't say that. I said sentiments are not the\\  same as arguments. The support of sentiments for the\\  sake of supporting sentiments is one of the stupider\\  things I've heard in my lifetime. Sorry to say. Not that I\\  think that reflects on you in anyway.”\end{tabular}                                                                                                                                                                                                                                                                                                                                                                                                \\  \hline
empathy/reciprocity         & \begin{tabular}[c]{@{}c@{}}The text recognizes another person's perspective \\ or situation, even if the author is not in that place.\end{tabular}                                                                                           & \begin{tabular}[c]{@{}c@{}} I agree, but then where that line is drawn maybe the \\ battle ground. Exactly, but we need to be honest in the discussion, \\ rather than using edge cases that distort  \\  the reality of the systems we are aiming to improve.”\end{tabular}                                                                                                                                                                                                                                                                                                                                                                                                                                                                                                   \\  \hline
emotional language          & \begin{tabular}[c]{@{}c@{}}The text contains (positive or negative) sentiment \\  related adjectives, swear words, offensive, \\ aggressive, or satirical language.\end{tabular}                                                                  & \begin{tabular}[c]{@{}c@{}}“What the fuck is this ceremony? Bowing every 5 steps?\\  Loyal to the Queen? Jesus Christ this is an antique\\  piece of shit we need to do away with.”\end{tabular}                                                                                                                                                                                                                                                                                                                                                                                                                                                                                                                                                                               \\  \hline
collective rhetoric         & \begin{tabular}[c]{@{}c@{}}The text refers to a collective\\  and specific uniting features of it.\end{tabular}                                                                                                                              & \begin{tabular}[c]{@{}c@{}}“My point was we the people are paying tax\\  revenue to deal with this, instead of collecting tax revenue.”\end{tabular}                                                                                                                                                                                                                                                                                                                                                                                                                                                                                                                                                                                                                         \\  \hline
general information         & \begin{tabular}[c]{@{}c@{}}The text mentions events or facts \\ without making a specific argument.\end{tabular}                                                                                                                             & \begin{tabular}[c]{@{}c@{}}“For the curious:\\ $14.22 in 1890 = $388.08 in 2015\\ $2,069.90 in 1890 = $56,489.79 in 2015”\end{tabular}                                                                                                                                                                                                                                                                                                                                                                                                                                                                                                                                                                                                                                         \\  \hline
we vs. them                       & \begin{tabular}[c]{@{}c@{}}The text either directly contradicts \\two social groups, taking the position of one side,\\ or mentions the importance of \\one group/or a group being good/bad in a competitive context.\end{tabular}                                                                                                                                                                                         & \begin{tabular}[c]{@{}c@{}}``AOC wants us to have a corrupt system \\like South America where the top 0.1\% still control everything \\though the government and the 9.9\% get fucked\\ out of everything they own. \\I'm in the top 10\% of income earners in the USA.\\ It isn't very hard to get into.\\ We are the ones who get fucked by these socialist policies.''\end{tabular}                                                                                                                                                                                                                                                                                                                                                                                                                                                                                                                                                                                                                                                                                                                                                            \\\hline
other                       & Other types of text not included in the categories.                                                                                                                                                                                          & {[}deleted{]}                                                                                                                                                                                                                                                                                                                                                                                                                                                                                                                                                                                                                                                                                                                                                                  \\  \hline
unsupported argument        & \begin{tabular}[c]{@{}c@{}}The text contains a statement or makes an \\ inference without providing evidence to support it.\end{tabular}                                                                                                     & \begin{tabular}[c]{@{}c@{}}“Not at all surprised. Isn't the conservative answer to\\  everything 'more guns, more violence'?”\end{tabular}                                                                                                                                                                                                                                                                                                                                                                                                                                                                                                                           
\end{tabular}}                                                                                                         
\end{table}

\begin{table}
\caption{List of subreddits for which we downloaded user comments}
\label{subreddits}
\begin{tabular}{llll}
\multicolumn{4}{c}{Subreddits}                                                      \\ \hline
EnoughTrumpSpam     & conservatives  & PoliticalDiscussion & conspiracy             \\
Anarcho\_Capitalism & censorship     & obama               & ukpolitics             \\
MensRights          & afghanistan    & AmericanPolitics    & raisedbynarcissists    \\
vegan               & PoliticalHumor & Sunlight            & trump                  \\
india               & atheism        & Libertarian         & Republican             \\
privacy             & AskALiberal    & progressive         & Anarchism              \\
occupywallstreet    & Economics      & democrats           & canada                 \\
tsa                 & Feminism       & iran                & EndlessWar             \\
JoeBiden            & anonymous      & uspolitics          & climateskeptics        \\
Conservative        & collapse       & moderatepolitics    & GreenParty             \\
politics            & racism         & Liberal             & Good\_Cop\_Free\_Donut \\
unpopularopinion    & antiwar        & 911truth            & Marxism                \\
GenderCritical      & union          & LGBTnews            & me\_irlgbt             \\
exmuslim            & humanrights    & alltheleft          &                    \\ \hline   
\end{tabular}
\end{table}

\begin{table*}
\caption{Political subreddits that changed their reaction mechanisms between 2010 and 2018}
\label{reactions}
\begin{tabular}{llll}
\textbf{subreddit} & \textbf{intervention} & \textbf{start month} & \textbf{end month} \\ \hline
Conservative       & up- \& downvote       & 2012-12          & 2013-10        \\
Conservative       & upvote                & 2013-11          & 2014-06       \\
Conservative       & no votes                & 2014-07          & 2018-04        \\
EnoughTrumpSpam    & up- \& downvote       & 2010-01          & 2016-01        \\
EnoughTrumpSpam    & upvote                & 2016-01         & 2018-04        \\
GenderCritical     & up- \& downvote       & 2013-09          & 2013-09        \\
GenderCritical     & upvote                & 2013-10          & 2014-06        \\
GenderCritical     & no votes                & 2014-07         & 2018-04        \\
atheism            & up- \& downvote       & 2010-01          & 2015-06        \\
atheism            & upvote                & 2015-06          & 2018-04        \\
exmuslim           & up- \& downvote       & 2010-01          & 2011-07        \\
exmuslim           & upvote                & 2011-08          & 2018-01        \\
politics           & up- \& downvote       & 2010-01         & 2014-01        \\
politics           & upvote                & 2014-01          & 2014-12        \\
politics           & no votes                & 2014-12         & 2018-04       \\
ukpolitics         & up- \& downvote       & 2010-01          & 2015-01        \\
ukpolitics         & upvote                & 2015-01          & 2018-04        \\
unpopularopinion   & up- \& downvote       & 2010-01          & 2018-02        \\
unpopularopinion   & no votes              & 2018-02          & 2018-04        \\
vegan              & up- \& downvote       & 2010-01          & 2017-11        \\
vegan              & no votes              & 2017-11          & 2018-04        \\ \hline
\end{tabular}
\end{table*}

\begin{table*}
\caption{Matched subreddits for controlling ``maturation'' effects in the DID models}
\label{treatment}
\begin{tabular}{cc}
\textbf{\begin{tabular}[c]{@{}l@{}}Treatment\\ subreddit\end{tabular}} & \textbf{\begin{tabular}[c]{@{}l@{}}Control\\ subreddit\end{tabular}} \\ \hline
Conservative                                                           & democrats                                                            \\
GenderCritical                                                         & MensRights                                                           \\
politics                                                               & Liberal                                                           \\
EnoughTrumpSpam                                                        & Good\_Cop\_Free\_Donut                                               \\
atheism                                                                & uspolitics                                                           \\
exmuslim                                                               & progressive                                                              \\
ukpolitics                                                             & AmericanPolitics                                                             \\
unpopularopinion                                                       & PoliticalHumor                                                       \\
vegan                                                                  & Republican   \\ \hline                                                       
\end{tabular}
\end{table*}
\clearpage 
\section{Confirmatory Factor Analysis}

\begin{figure*}[!h]
    \centering
    \includegraphics[width=0.8\textwidth]{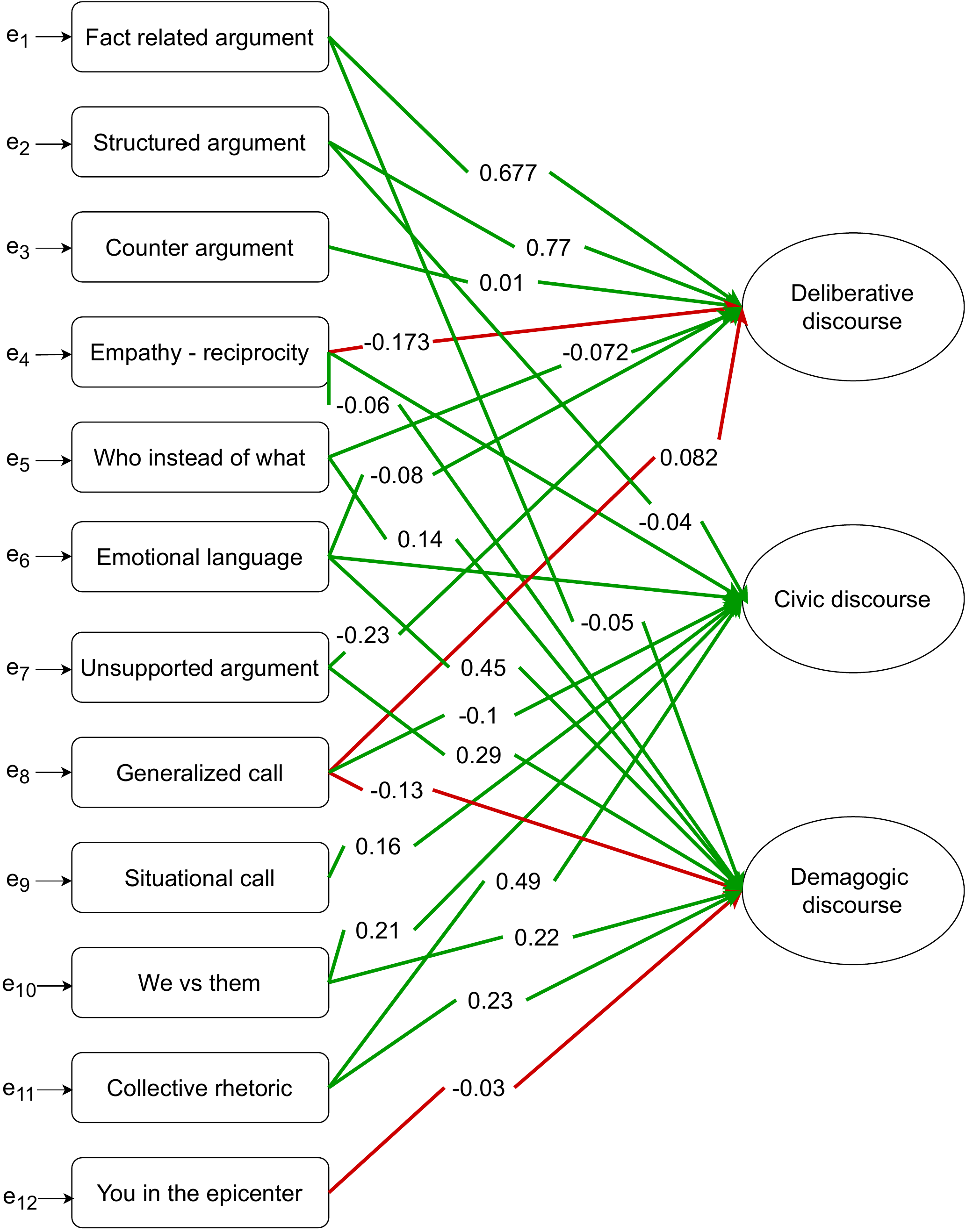}
    \caption{Confirmatory factor analysis results for the expansive model. The arrows define an existing connection between item and factor. With green color we present an association that complied with the theoretic conceptualization of political theories, while with red color we present contradictory results.}
    \label{fig:expansive}
\end{figure*}

\begin{figure*}[!h]
    \centering
    \includegraphics[width=0.9\textwidth]{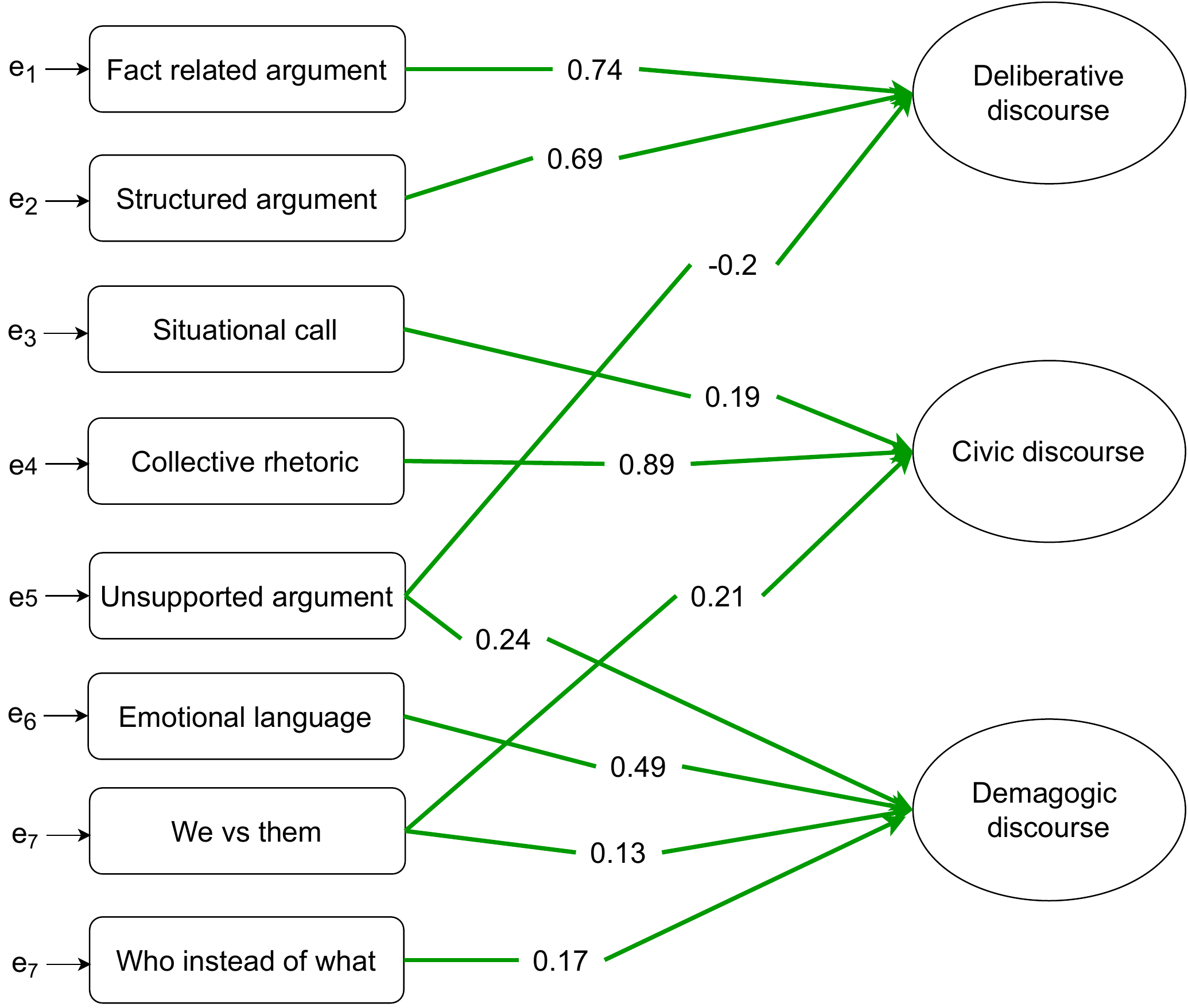}
    \caption{Confirmatory factor analysis results for the robust model. The arrows define an existing connection between item and factor. With green color we present an association that complied with the theoretic conceptualization of political theories, while with red color we present contradictory results.}
    \label{fig:robust}
\end{figure*}

\clearpage
\section{Difference-in-Difference Diagnostic Plots} \label{did_appendix}

\begin{figure*}[!h]
    \centering
    \includegraphics[width=0.85\textwidth]{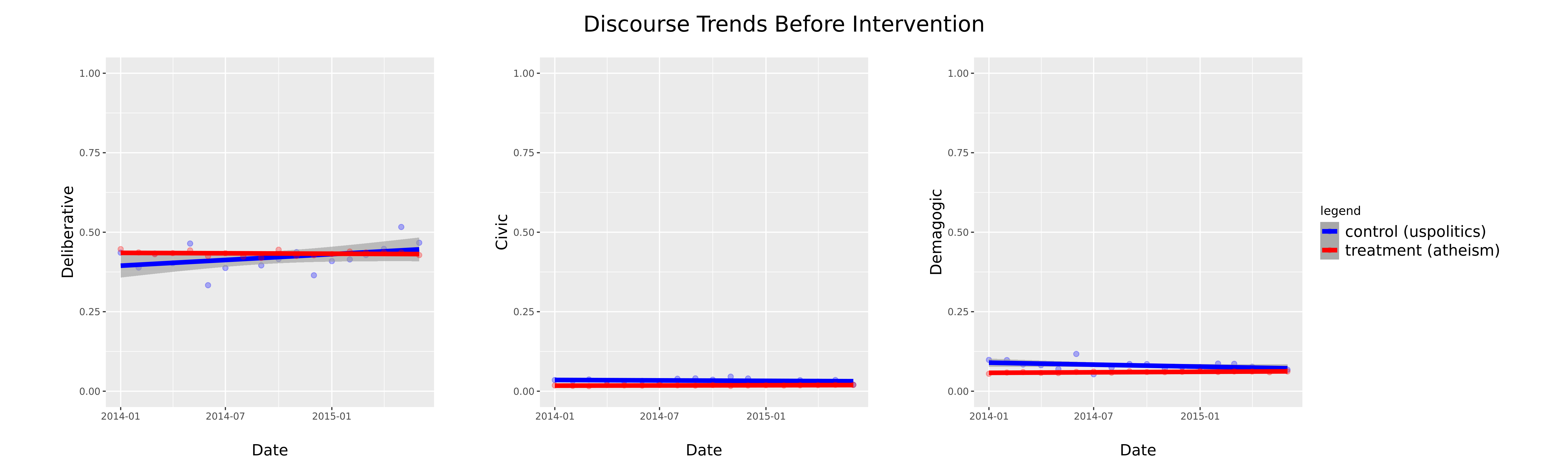}
    \caption{Comparison of  discourse dynamics for the \textit{atheism} (treatment) and \textit{uspolitics} (control) subreddits prior to a change  from  up/downvotes to only upvotes for the treatment subreddit. The shadows represent the confidence intervals of the fitted regression lines. As it is visible, the trends between treatment and control either match perfectly, or there is a small deviation. Therefore, we assume that the ``parallel trends'' assumption is fulfilled. }
    \label{fig:test_plot1}
\end{figure*}
\begin{figure*}[!h]
    \centering
    \includegraphics[width=0.85\textwidth]{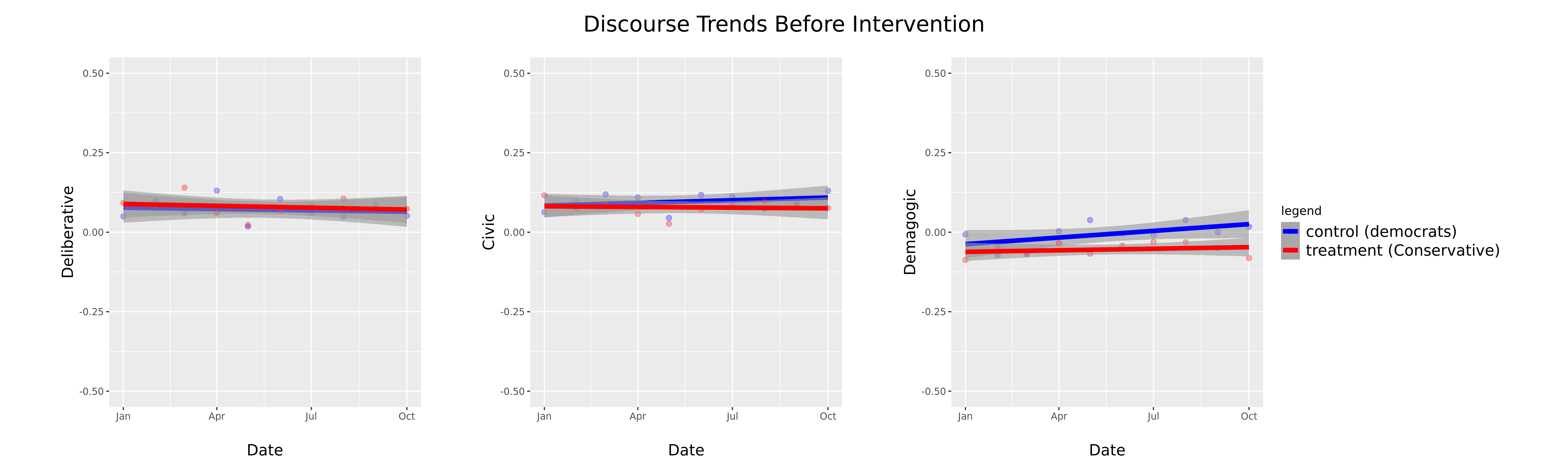}
        \caption{Comparison of  discourse dynamics for the \textit{Conservative} (treatment) and \textit{democrats} (control) subreddits prior to a change  from  up/downvotes to only upvotes for the treatment subreddit.The shadows represent the confidence intervals of the fitted regression lines. As it is visible, the trends between treatment and control either match perfectly, or there is a small deviation. Therefore, we assume that the ``parallel trends'' assumption is fulfilled.}
    \label{fig:test_plot2}
\end{figure*}
\begin{figure*}[!h]
    \centering
    \includegraphics[width=0.85\textwidth]{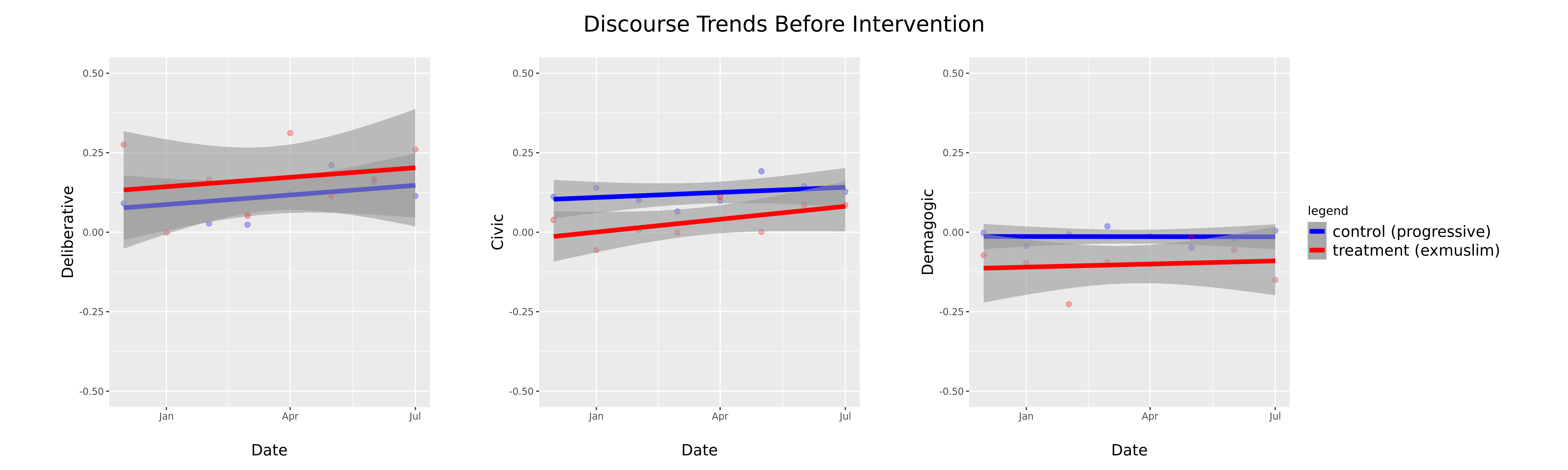}
        \caption{Comparison of  discourse dynamics for the \textit{exmuslim} (treatment) and \textit{progressive} (control) subreddits prior to a change  from  up/downvotes to only upvotes for the treatment subreddit.The shadows represent the confidence intervals of the fitted regression lines. As it is visible, the trends between treatment and control either match perfectly, or there is a small deviation. Therefore, we assume that the ``parallel trends'' assumption is fulfilled.}
    \label{fig:test_plo3t}
\end{figure*}
\begin{figure*}[!h]
    \centering
    \includegraphics[width=0.85\textwidth]{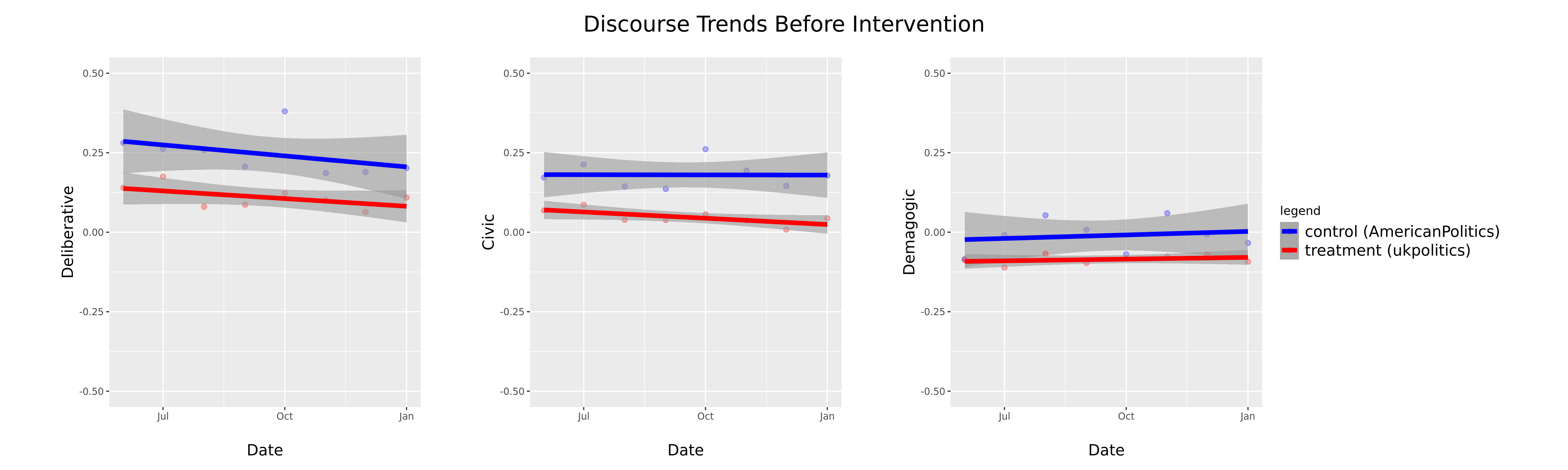}
        \caption{Comparison of  discourse dynamics for the \textit{ukpolitics} (treatment) and \textit{AmericanPolitics} (control) subreddits prior to a change from  up/downvotes to only upvotes for the treatment subreddit.The shadows represent the confidence intervals of the fitted regression lines. As it is visible, the trends between treatment and control either match perfectly, or there is a small deviation. Therefore, we assume that the ``parallel trends'' assumption is fulfilled.}
    \label{fig:test_plot4}
\end{figure*}
\begin{figure*}[!h]
    \centering
    \includegraphics[width=0.85\textwidth]{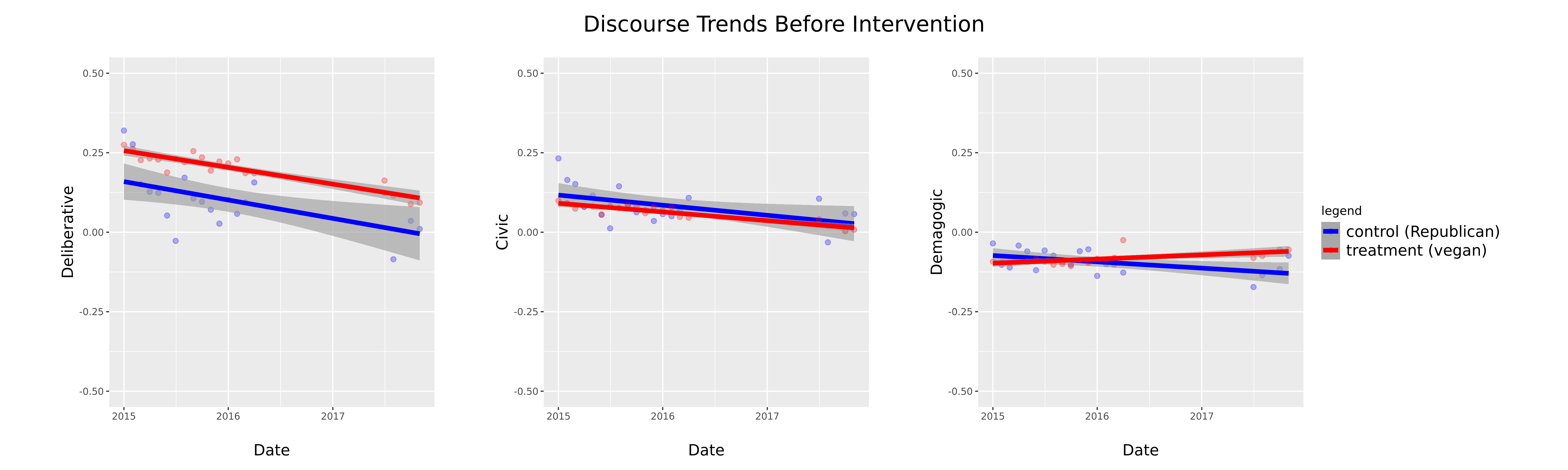}
        \caption{Comparison of  discourse dynamics for the \textit{vegan} (treatment) and \textit{Republican} (control) subreddits prior to a change from  up/downvotes to only upvotes for the treatment subreddit.The shadows represent the confidence intervals of the fitted regression lines. As it is visible, the trends between treatment and control either match perfectly, or there is a small deviation. Therefore, we assume that the ``parallel trends'' assumption is fulfilled.}
    \label{fig:test_plot5}
\end{figure*}
\begin{figure*}[!h]
    \centering
    \includegraphics[width=0.85\textwidth]{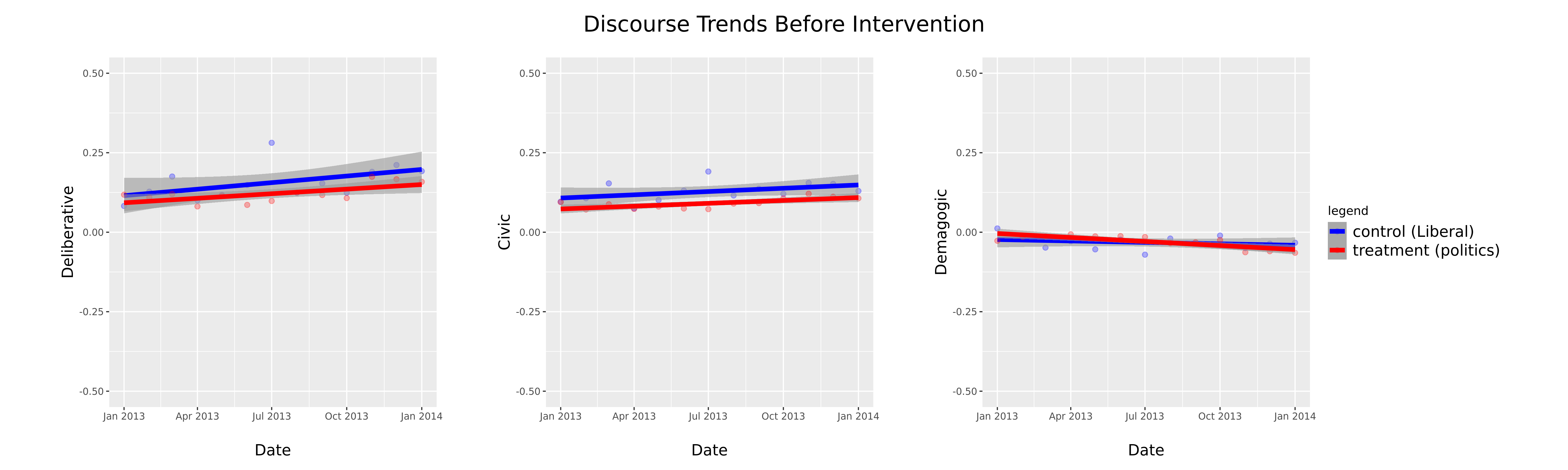}
        \caption{Comparison of  discourse dynamics for the \textit{politics} (treatment) and \textit{Liberal} (control) subreddits prior to a change  from  up/downvotes to only upvotes for the treatment subreddit.The shadows represent the confidence intervals of the fitted regression lines. As it is visible, the trends between treatment and control either match perfectly, or there is a small deviation. Therefore, we assume that the ``parallel trends'' assumption is fulfilled.}
    \label{fig:test_plot6}
\end{figure*}
\end{document}